\documentclass[aps,twocolumn,longbibliography,prb,byrevtex]{revtex4-1}
\usepackage{graphicx}    
\usepackage{dcolumn}    
\usepackage{bm}             
\usepackage{amssymb}   
\usepackage{amsmath}    
\usepackage{subfigure}   
\usepackage{color}                     

\begin{document}
\title{Parafermions, induced edge states and domain walls in the fractional quantum Hall effect spin transitions}
\author{Jingcheng Liang}
\thanks{Now at the Department of Physis, University of California at San Diego, 9500 Gilman Drive, La Jolla, CA 92093 } 
\author{George Simion}
\thanks{Now at IMEC , Kapeldreef 75, 3001, Leuven, Belgium} 
\author{Yuli Lyanda-Geller}
\email{yuli@purdue.edu} \affiliation{Department of Physics and Astronomy and Purdue Quantum Insitute, Purdue University, West Lafayette IN, 47907 USA}

\begin{abstract}
Search for parafermions and Fibonacci anyons, which are excitations obeying non-Abelian statistics, is driven both by the quest for deeper understanding of nature and prospects for universal topological quantum computation. However, physical systems that can host these exotic excitations are rare and hard to realize in experiments. Here we study the domain walls and the edge states formed in spin transitions in the fractional quantum Hall effect. Effective theory approach and exact diagonalization in a disk and torus geometries proves the existence of the counter-propagating edge modes with opposite spin polarizations at the boundary between the two neighboring regions of the two-dimensional electron liquid in spin-polarized and spin-unpolarized phases. By analytical and numerical analysis, we argue that these systems can host parafermions when coupled to an s-wave superconductor and are experimentally feasible. We investigate settings based on $\nu=\frac{2}{3}$, $\nu=\frac{4}{3}$ and $\nu=\frac{5}{3}$ spin transitions and analyze spin-flipping interactions that hybridize counter-propagating modes. Finally, we discuss spin-orbit interactions of composite fermions. 
\end{abstract}

\maketitle

\section{I. Introduction}
Since early years of quantum physics, it has been recognized that symmetry with respect to an exchange of particles results in the two possible quantum statistics, Bose-Einstein statistics for particles with integer spin and Fermi statistics for particles with half-integer spin. About fifty years later researchers  realized that particles/quasiparticles confined to one dimension \cite{Su} or two dimensions \cite{Leinaas,Wilczek} can obey a different statistics, which F. Wilczek called anyon statistics, for which an exchange results in the wavefunction picking up any possible quantum-machanical phase. Furthermore, it was realized \cite{Frohlich,ReadMoore,wen1991nonabelian,ReadGreen,Ivanov} that for degenerate states, an exchange, or more appropriately, braiding of particles or quasiparticles in two dimensions may result in nontrivial unitary transformation of the corresponding wavefunctions, i.e. in the non-abelian statistics. It was recognized recently that the non-abelian statistics opens new ways of approaching a fault-tolerant quantum computation.
An approach to the topological quantum computation based on the Majorana fermions has been widely studied in recent years\cite{kitaev2001unpaired,nayak2008non,sau2010generic,alicea2011non}. However, it became apparent that such systems are not computationally universal because braiding operations for Majorana fermions cannot approximate all unitary quantum gates\cite{freedman2002two,baraban2010resources}. In order to realize the universal topological quantum computation, other kinds of non-Abelian anyons are required. In particular, parafermions have been shown to have denser rotation groups and their braiding operations can enable two-qubit entangling gates\cite{fendley2012parafermionic,alicea2016topological}. Furthermore, a two dimensional array of parafermions can support Fibonacci anyons with universal
braiding statistics\cite{mong2014universal}. Therefore, it is of great interest to find experimentally realizable systems which can host parafermions. In a seminal paper\cite{clarke2013exotic}, Clarke, Allicea and Shtengel proposed that parafermions can appear in the fractional quantum Hall effect  (FQHE) regime at filling factors $\nu=\frac{1}{m}$  if two counter-propagating edge states from two adjacent 2D electron gases with opposite g-factors are gapped by the proximity superconducting pairing and spin-orbit induced tunneling.

 Here we propose that a single layer of the 2D electron gas in a magnetic field near the spin transition between the filling factor  $\nu=\frac{2}{3}$ spin-unpolarized  and spin-polarized states can be used to create a domain wall that will host parafermions when coupled to an s-wave superconductor, and use exact diagonalization of small systems in order to confirm this result microscopically. We discuss feasible experimental settings, analyze viable spin-flipping mechanisms capable of gapping counterpropagating modes with oppoosite spin, and discuss possible realizations of topological superconductivity and parafermions in spin transitions besides $\nu=\frac{2}{3}$. 

\begin{figure}
\centering
\subfigure
{
\begin{minipage}[b]{0.22\textwidth}
\includegraphics[width=4.1cm,height=4.1cm]{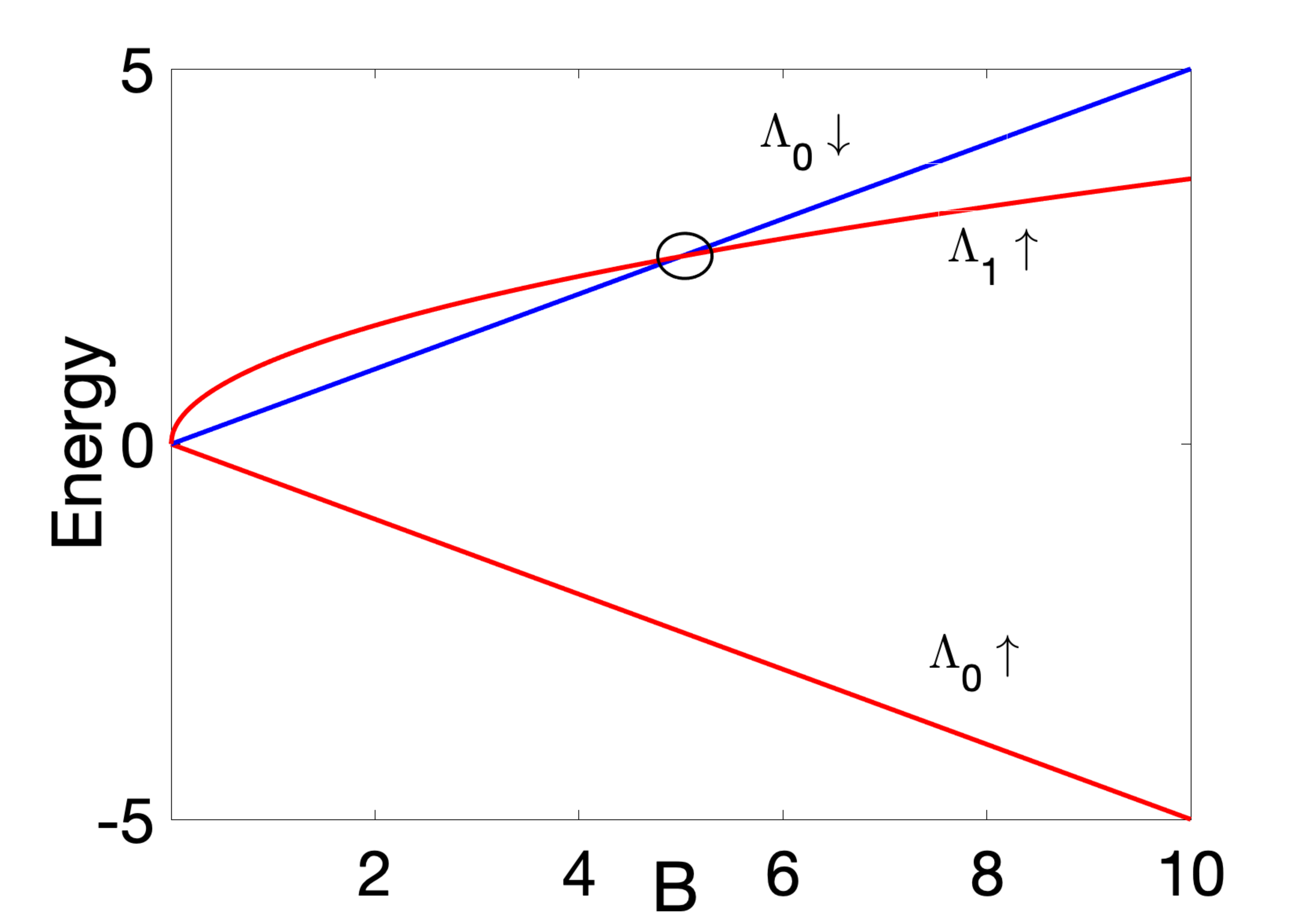}
\put(-120,100){(a)}
\end{minipage}

\label{1a}
}
\subfigure
{
\begin{minipage}[b]{0.22\textwidth}
\includegraphics[width=4.1cm,height=4.1cm]{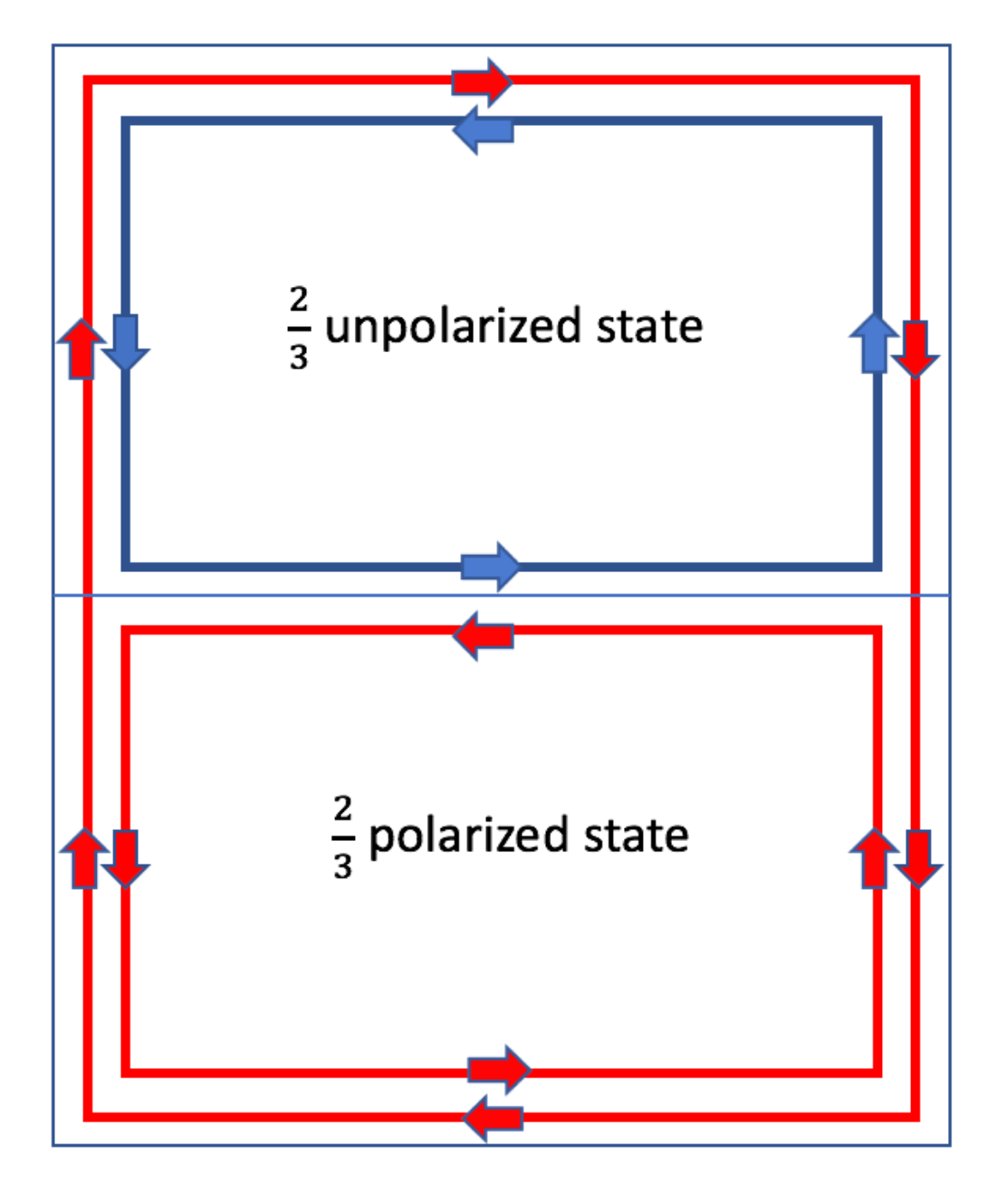}
\put(-124,100){(b)}
\end{minipage}
\label{1b}
}
\vspace{2mm}
\caption{(Color online). a: The schematic plot of the composite fermion energy levels. When the magnetic field is increased, there is a level crossing(black circle) of the $\Lambda_{0\downarrow}$ and $\Lambda_{1\uparrow}$ which leads to a spin transition from spin-unpolarized state to spin-polarized state. b: A schematic plot of the edge states. The arrows represent the direction of the their velocities and colors represent spin up(red) spin down(blue). \label{fig1}}
\end{figure}

The FQHE spin transitions  have been observed at the filling factor $\nu=\frac{2}{3}$ and other fractions\cite{eisenstein1990evidence,smet2001ising}, e.g. when an in-plane component of a tilted magnetic field is varied, resulting in a change in electron spin system. Such spin transition can be understood in terms of the composite fermion (CF) picture \cite{Jain}. The FQHE states at a filling factor $\nu=\frac{n}{2n-1}$ can be mapped onto the integer quantum Hall states at a filling factor $n$ of CFs. The energy of the n-th CF level is:
\begin{equation}
\label{e1}
E_{ns}=\hbar \omega_c^{cf}(n+\frac{1}{2})+sg\mu_BB,
\end{equation}
where the CF cyclotron energy $\hbar \omega_c^{cf}$  is proportional to the characteristic electron-electron interaction energy scale $\frac{e^2}{l_m}$,  $l_m =\sqrt{\hbar c}{e B_{\perp}}$ is the magnetic length, and $B_{\perp}$ is the out of plane component of the magnetic field $\mathbf{B}$. The second term is the Zeeman energy, the index $s=\pm1$, represents the up and down spin states of the composite fermions. Since the cyclotron and Zeeman energies have different magnetic field dependences, the levels $\Lambda_{p,\downarrow}$ and $\Lambda_{p+1,\uparrow}$ have to cross at some $B^*>0$, as shown in Fig.\ref{1a}. Therefore, at $B<B^*$ electrons of $\frac{2}{3}$ state occupy $\Lambda_{0\uparrow}$ and 
$\Lambda_{0\downarrow}$ and it is in the spin-unpolarized phase, while at $B>B^*$ they occupy $\Lambda_{0\uparrow}$ and $\Lambda_{1\uparrow}$ states and it is in the spin-polarized phase. Furthermore, it has been shown that electrostatic gates control electron-electron interactions, so that in a triangular quantum well  $\hbar \omega_c^{cf}\propto\frac{e^2}{\sqrt{l_m^2+z_0^2}}$, where $z_0$ is the extent of the electron wavefunction in the direction of the spatial quantization.  Therefore it is possible to induce the
spin-polarized and spin-unpolarized fractional quantum Hall phases in a single quantum well underneath different electrostatic gates. In this case a controlled domain wall that separates regions with different spin polarizations should emerge\cite{kazakov2017mesoscopic,simion2017disorder}.   Experimentally transitions can then be achieved by both tuning the effective Coulomb interaction  and/or by tuning the Zeeman coupling via the in-plane component of the magnetic field \cite{wu2017parafermion}.

The boundary berween polarized and unpolarized regions of the 2d electron liquid results in edge-like states, which we will call edge states despite they are, strictly speaking, different from the true edge states flowing at the boundary of the quantum Hall samples. The reason for a difference between these two kinds of edge states is obvious. For edges at the sample boundary, the spin-unpolarized or spin-polarized phase has to decrease its density (filling factor) from $\nu=\frac{2}{3}$ in the bulk to $\nu=0$ at the sample boundary. For the boundary between two $\nu=\frac{2}{3}$ phases, the density stays nearly constant, the change in density has to be less than corresponding to the width of the $\nu=\frac{2}{3}$ plateau. 

The first goal of our paper is to demonstrate the existence and investigate the nature of the edge states flowing through the domain wall between polarized and unpolarized fractional quantum Hall spin regions.
 The edge states of the quantum Hall systems were widely studied over the years\cite{macdonald1990edge,beenakker1990edge,macdonald1991edge,wen1991topological,wen1992theory,chamon1994sharp,chklovskii1995structure,wen1995topological,Levitov,Fradkin}. For the filling factor $\nu=\frac{2}{3}$, edge states of the fractional quantum Hall liquid at the sample boundaries have been studied in both spin-polarized and various kinds of unpolarized phases\cite{mcdonald1996topological,moore1997edge,Rezayi,wu2012microscopic,wang2013edge}. In has been predicted that both phases of $\nu=\frac{2}{3}$ electron liquid are characterized by two counterporopagating  edge modes at the sample boundaries. 

A new setting emerges on the spatial boundary between the two different topological orders, i.e. the domain wall between spin-polarized and unpolarized phases realized in the neighboring regions of the 2D electron liquid. When the $\frac{2}{3}$ polarized and unpolarized states are far apart, there are four edge modes with two states moving in one direction and two states moving in the opposite direction. When they are brought closer together, one can anticipate that there are only two edge modes are left, see Fig.\ref{1b}. This can be understood intuitively in terms of the CF picture. The edge states associated with the common $\Lambda_{0\uparrow}$ level will merge and disappear, and only the edge states associated with the $\Lambda_{0\downarrow}$ and $\Lambda_{1\uparrow}$ levels will remain. This picture implies that the remaining edge states propagate in opposite directions and carry opposite spins. We will present simple qualitative ariguments to justify this picture, and demonstrate it rigorously by using the effective field theory. We will also apply exact diagonalization method in a disk and torus geometries. 

 It then follows that the domain wall excitations in the proximity of an s-superconductor are possibly characterized by a parafermion non-abelian statistics. The domain wall system is similar to the setting discussed in \cite{clarke2013exotic} with a boundary of two fractional Hall liquids having opposite values of the electron g-factor at a filling factor $\nu=1/3$ in proximity to an s-superconductor. We will see indeed that the proximity coupling of the fractional quantum Hall ferromagnet domain wall area to an s-wave superconductor induces parafermions. 

In an experimental setting, one needs to control the onset of topological supercondutivity and be able to induce and move the boundaries between topological and non-topological superconducting regions. Parafermions must be located at the boundaries between these regions. Thus an ability to change the boundaries between these regions allows to move parafermions, which would be ultimately necessary for their braiding. However, when crossing between the composite fermion Landau levels takes place, proximity coupling to the domain wall will always yield a topological proximity superconductivity,  in much the same way as it happens in topological insulators \cite{Kane,Alicea2012}. Then no trivial proximity superconducting at any value of induced superconducting order parameter is expected. In order to have both types of proximity superconductivity and potentially a boundary between the two regions with different superconducting order that differs from the location of  superconducting contacts , one has to induce a gap due to tunneling between the two counterpropagating edge states with opposite spin. When this tunneling is tuned, an onset of topological superconductivity depends on the competition between tunneling and proximity superconducting gaps. If an induced superconducting order parameter in the domain wall changes its amplitude along the domain wall, the possibility to move parafermions along the domain emerges.

We will study the effective theory of states in the domain wall when two gapping mechanisms are introduced: superconducting pairing and spin-flip tunneling across the domain wall caused by an in-plane component of a  magnetic field, and demonstrate an emergence of parafermions in this system. We find that in a single quantum well the spin-orbit coupling is negligible between fractional quantum Hall edge states with opposite spins belonging to the first electron Landau level. Our simulation shows that despite a small g-factor in a GaAs system, gapping of the edge states can be realized by an in-plane component of a the magnetic field. We also propose that a good candidate for observing parafemions is CdMnTe system with the effective Zeemann splitting of the order to the cyclotron frequency, where the fractional quantum Hall effect has been observed in \cite{Weiss}. Furthermore, in this system, the fractional quantum Hall effect spin transition was observed at $\nu=5/3$, and there have been also signatures of spin transition at  $\nu=4/3$ . In these cases, the fractional quantum Hall edge states that potentially experience crossing originate from different electron Landau levels, and in these circumstances spin-orbit interactions result in sizable anticrossing gap. Because spin-orbit coupling can be effectively tuned by electrostatic gates, this setting would allow to tune the parafermion zero modes and their braiding using only gate voltage.  

We would like to underscore that our results not only provide a path to a new platform realizing universal topologic al quantum computation, but also illustrate a general method to study the edge states on the boundary of systems with different topological orders.We also developed a scheme for numerical modeling of fractional quantum Hall states in proximity of an s-superconductor.

Our paper is organized as follows: 
In Sec.II we analyse edge states on the boundary between topologically distinct $\frac{2}{3}$ spin polarized and unpolarized states. Sec III is devoted to the numerical calculations of edge states on on the disk, and Sec. IV presents the numerical calculations of edge modes on the torus. In Sec.V we disuss the emergence of parafermion zero modes and Sec. VI describes the numerical calcultion of these modes. In a brief Sec. VII we will discuss a possible parafermion setting based on $\nu=\frac{4}{3}$ and $\nu=\frac{5}{3}$  spin transitions and spin-orbit  interactions of composite fermions. We summarize our results in Sec. VIII. In Appendix, we evaluate spin-flipping interactions of the quasiparticles orignating from the lowest Landau level.

\section{II. Analytic consideration of edge states on the boundary between $\nu=\frac{2}{3}$ spin-polarized and spin-unpolarized fractional Quantum Hall states.}
In this section, we will use the effective theory in order to analyze the structure of the edge states on the boundary between $\frac{2}{3}$ spin polarized and unpolarized phases analytically. We will quantitatively show that there remain only two edge modes, which propagate in opposite directions and have carry opposite spins. An analytic theory is essential because not only it sheds light on the numerical calculations in the following sections, but is also necessary for the study of the emergence of parafermions. 

\begin{figure}

\subfigure
{
\begin{minipage}[b]{0.22\textwidth}
\includegraphics[width=4.5cm,height=4cm]{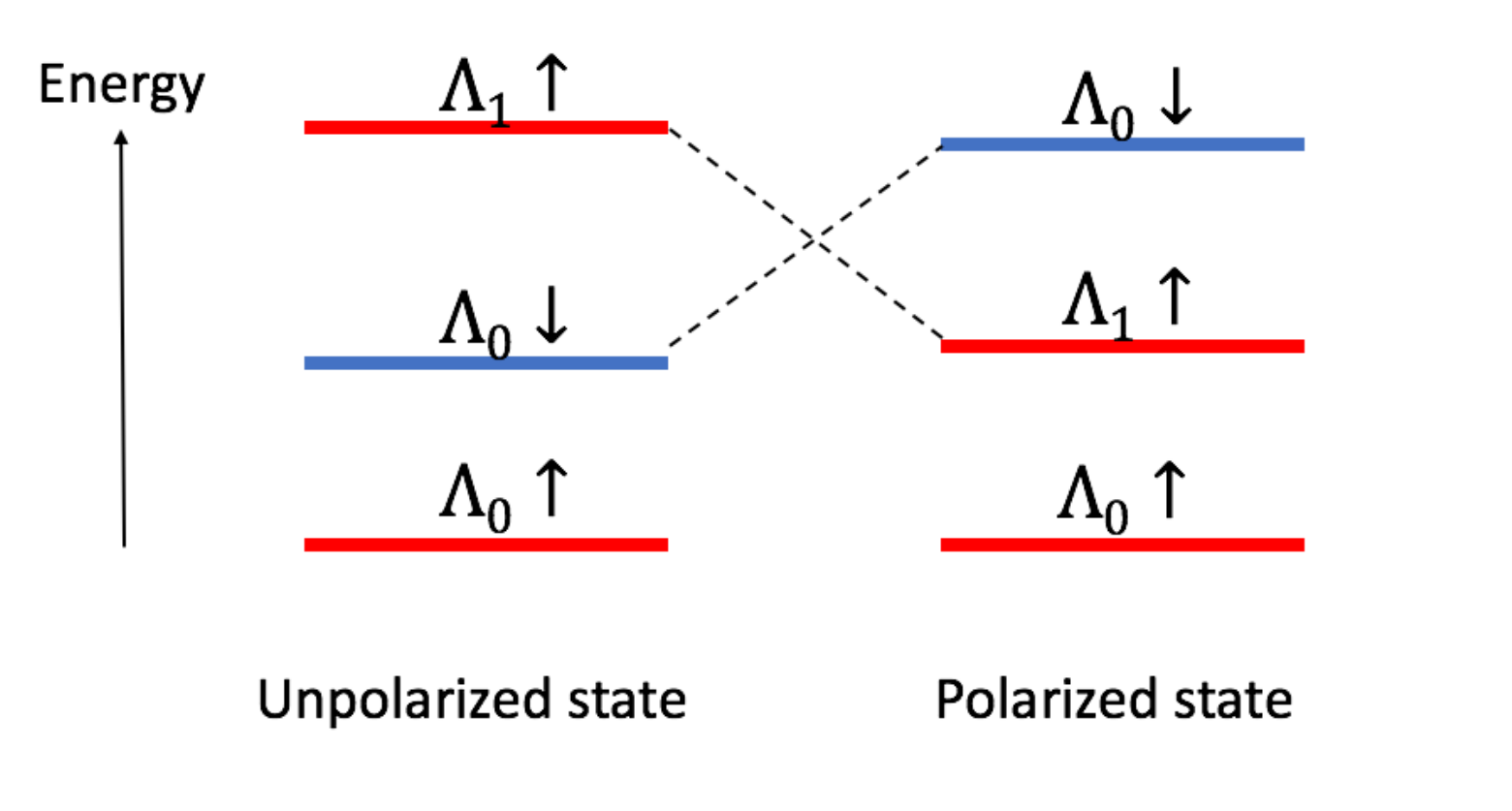}
\put(-130,110){(a)}
\end{minipage}

\label{2a}
}
\hfill
\subfigure
{
\begin{minipage}[b]{0.22\textwidth}
\includegraphics[width=4cm,height=4cm]{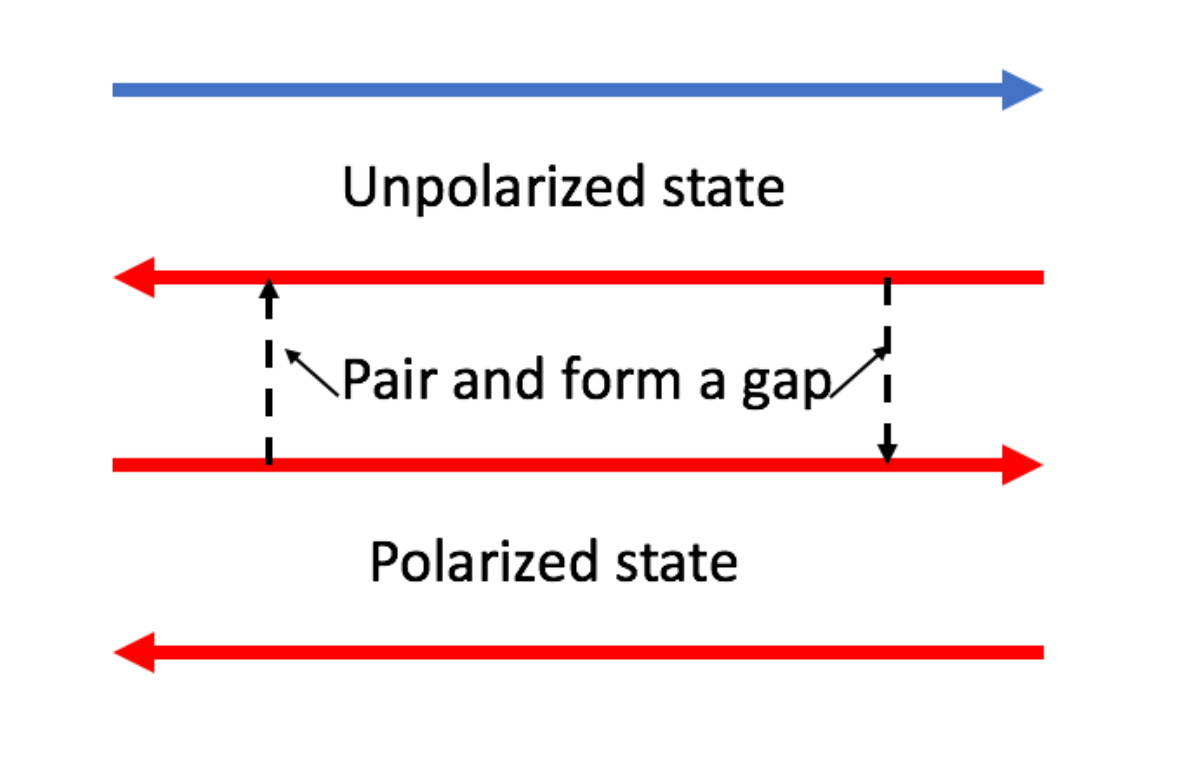}
\put(-120,110){(b)}
\end{minipage}
\label{2b}
}
\subfigure
{
\begin{minipage}[b]{0.45\textwidth}
\includegraphics[width=8cm,height=4cm]{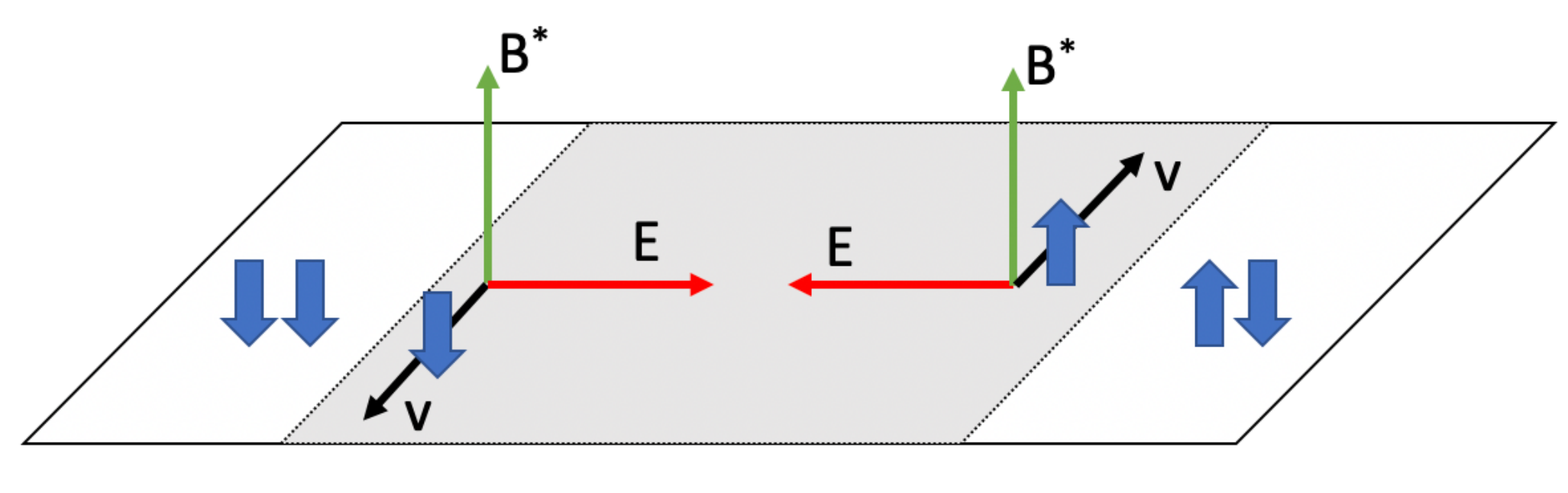}
\put(-120,110){(b)}
\end{minipage}
\label{2c}
}
\caption{(Color online). a: A schematic plot of the composite fermion energy levels in the bulk of the spin-polarized and spin-unpolarized regions. Potential barriers for composite fermions of the $\Lambda_{1\uparrow}$   Landau level  and  composite fermions of the $\Lambda_{0\downarrow}$ Landau level are shown.  b: Possible edge modes on the boundary of spin-polarized and spin-unpolarized regions. The two modes in the middle corresponding to $\Lambda_{0\uparrow}$ can pair and form a gap. They are not considered in the low energy theory. c: illustration of the crossed effective magnetic field ${\mathbf B}^{*}$ and electric field ${\mathbf E}$ resulting in counterpropagating modes of opposite spin and opposite velocities ${\mathbf v}$ in the shaded domain wall area. The electric field 
 ${\mathbf E}$ is opposite for two u and down spin states, as follows from potential barriers in Fig.~1a. \label{fig2}}
\end{figure} 
Before considering the quantitative theory, we first qualitatively explain why there are edge states on the boundary between the two regions with topologically different orders. The formation of the edge modes are always related to the confinement potentials acting at the edges. Naively it seems that there is no confinement potential around the internal edge in our case. However, we actually have intrinsic spin-dependent confinement potential. Indeed, from the analysis in the introduction, we see that there is a level crossing between the edge states with opposite spins. The composite fermions in the $\Lambda_{0\downarrow}$ level in the spin-unpolarized phase can not tunnel into region of the spin-polarized phase  because $\Lambda_{0\downarrow}$ level has a higher energy there. The same story also applies for the composite fermions of the $\Lambda_{1\uparrow}$   Landau level  in the spin polarized phase, see Fig.\ref{2a}. Therefore, the composite fermions of these two levels characterizing opposite spins are subject to the effective spin-dependent potential confinements. A spatial gradient of this spin-dependent potential constitutes a spin-dependent electric field.
Depending on the experimental setting, the spin-dependent confinement and electric fields are controlled either by varying $z_0$ extent of the electron wavefunction by electrostatic gates or by a spatial variation of the Zeeman coupling of the electron spin to an external magnetic field.  The spin-dependent electric field acts together with the 
effective residual magnetic field, which is due to joint effect of the external magnetic field and the Chern-Simons field. For $\nu=\frac{2}{3}$, the effective magnetic field is negative \cite{wu2012microscopic}, but it is the same in both the spin-polarized and unpolarized phases.
 Assuming a simple model, in which the change of potentials for each of the spins is linear in coordinate across the domain wall, we find that composite fermions with spin polarization up are subject to the crossed effective magnetic field and the electric field of one sign. For the spin polarization  down, we again have crossed magnetic and electric fields, with an effective magnetic field being the same as for the spin up, but with the opposite electric field. Thus, we have two counterpropagating composite fermion edge states of opposite spins due to these crossed  effective magnetic and electric fields. This situation is analogous to the case of edge states at the domain wall between polarized and unpolarzed 2D regions in the case of the integer Quantum Hall effect \cite{simion2017disorder}. If Hamiltonian of the system contains only the out of plane magnetic field, the counterpropagating edge states experience no backscattering. An in-plane magnetic field results in hybridization of these edge states with opposite spins.

We now discuss the effective theory for the boundary of the polarized and unpolarized regions. The Lagrangian density in the effective theory for the bulk fractional quantum Hall state can be written in the form\cite{wen1992theory,wen1995topological}:
\begin{eqnarray}
\label{e2}
\mathcal{L}&=&-\frac{1}{4\pi}K_{II'}a_{I\mu}\partial_\nu a_{I'lambda}\epsilon^{\mu\nu\lambda}-
\frac{e}{2\pi}q_IA_\mu\partial_\nu a_{I\lambda}\epsilon^{\mu\nu\lambda}\nonumber\\&+&s_I\omega_\mu\partial_\nu 
a_{I\lambda}\epsilon^{\mu\nu\lambda},
\end{eqnarray}
where $a_{I\mu}$ represents $n$ Abelian Chern Simons (CS) gauge fields, $A_\mu$ is the electromagnetic gauge field, $\omega_\mu$ describes the curvature of the space, $\mathbf{K}$ is an $n\times n$ nonsingular integer matrix describing the coupling between the CS gauge fields, $\mathbf{ q}$ is an n-component integer vector describing the coupling between the CS gauge fields and the electromagnetic gauge field, $\mathbf{s}$ is an n-component half-integer vector describing the coupling between the CS gauge fields and the curvature. An Abelian quantum Hall state can be classified by a set $\{\mathbf{K},\mathbf{q},\mathbf{s}\}$, which determine the long distance properties of the state. Therefore, this set characterizes the topological order of the Abelian quantum Hall fluid. For the $\nu=\frac{2}{3}$ case, it takes the following values in the spin- polarized phase:
\begin{equation}
\label{e3}
\mathbf{K}_1=\begin{pmatrix}
      1&  2  \\
     2 &  1
\end{pmatrix}, \mathbf{q}_1=\begin{pmatrix}
         1 \\
        1
\end{pmatrix},\mathbf{s}_1=\begin{pmatrix}
      \frac{1}{2}   \\
      -\frac{1}{2}
\end{pmatrix}.
\end{equation}
For the spin-unpolarized phase this set is defined by:
\begin{equation}
\label{e4}
\mathbf{K}_2=\begin{pmatrix}
      1&  2  \\
     2 &  1
\end{pmatrix}, \mathbf{q}_2=\begin{pmatrix}
         1 \\
        1
\end{pmatrix},\mathbf{s}_2=\begin{pmatrix}
      \frac{1}{2}   \\
      \frac{1}{2}
\end{pmatrix}.
\end{equation}
The form of the $\mathbf{K} $ matrix can be understood in terms of the CF picture \cite{wu2012microscopic}. For the $\nu=\frac{2}{3}$ state, there are two components each occupying a single CF Landau level in a effective antiparallel magnetic field. Thus the corresponding contribution to the $\mathbf{K} $ matrix is $K_{ij}=-\delta_{ij}$. Each component should have two fluxes attached, so we add an integer 2 to each element of the $\mathbf{K} $ matrix. From Eqs. (\ref{e3}) and (\ref{e4}), we see that the only difference between the spin-polarized and unpolarized phases is the spin vector $\mathbf{s}$, as expected. In Eq. (\ref{e2}) the second and the third terms are similar. If we regard $\mathbf{q}$ as a vector describing the unit of the electric charge carried by the two CF components, $\mathbf{s} $ can be regarded as describing the ``curvature charge" carried by the CF components. 

We now consider the physics of the edge states. We note that for the edge states in the domain wall, due to absence of boundary with vacuum, simple arguments based on $\nu=1$ forward-lowing mode bordering vacuum and 1/3 backward-flowing mode of holes can no longer be applied even for spin-polarized phase. Similarly, analogous CF picture with two edges, one separating 2 and 1 filled CF Landau and the other separating 1 and 0, which upon antiparallel flux attachment become 2/3, 1/3 and 0, also does not work, as instead of vacuum we have a phase of nearly the same density, within the interval of densities on the $\nu=\frac{2}{3}$ plateau, at the boundary. Therefore the K-matrix description of Wen \cite{wen1995topological} is the reliable way to approach the solution of this problem.   

When the FQH liquid is confined by the boundaries of the sample, the action $S=\int dxdydt\mathcal{L}$, where $\mathcal{L}$ is given by Eq. (\ref{e2}), is not gauge invariant for the CS gauge fields. To restore the gauge invariance, one has to introduce an action that describes the edge physics:
\begin{equation}
\label{e5}
\mathcal{S}_{edge}=\frac{1}{4\pi}\int dt dx[\mathbf{K}_{IJ}\partial_t\phi_I\partial_x\phi_J-
\mathbf{V}_{IJ}\partial_x\phi_I\partial_x\phi_J].
\end{equation}

In the equation (\ref{e5}), we assume that the edge is along the $x-$axis,$\phi_I$ is the field that describes the $\textit{I}$th component of the edge branches, $a_{Ii}=\partial_i\phi_I$ and $\rho_I=\frac{1}{2\pi}\partial_x\phi_I$ is the density of the i-th branch. $\mathbf{K}$ is the same matrix as in the bulk phase, and one can show that its positive eigenvalues correspond to the left-moving branches, and its negative eigenvalues scorrespond to the right-moving branches. $\mathbf{V}$ is a positive-definite matrix that encodes the non-universal interactions between edge branches. We would like to study now the properties of the edge states between two Abelian FQH phases. Assuming the edge is along $y=0$ axis, and using the same gauge argument as in Ref. \cite{wen1995topological}, we find that the $\{\mathbf{K},\mathbf{q},\mathbf{s}\}$ for the new state is 
\begin{equation}
\label{e6}
\mathbf{K}=\mathbf{K}_1\bigoplus -\mathbf{K}_2, \mathbf{q}=\mathbf{q}_1\bigoplus\mathbf{q}_2, \mathbf{s}=\mathbf{s}_1\bigoplus\mathbf{s}_2.
\end{equation}
A similar situation for bilayers was considered in Ref. \cite{mcdonald1996topological}, however the spin vectors did not play any role there. Here we include  spin vectors into the picture. In Eq. (\ref{e6}), $dim(K)=dim(K_1)+dim(K_2)$, and all edge branches are retained. After considering the tunneling perturbation, we see that two of the edge branches can be removed from the low energy theory. Following Ref. \cite{mcdonald1996topological}, we define the following quantities: 
\begin{eqnarray}
\label{e7}
\phi(m)=m_i\phi_i, \quad q(m)=m_iq_i, \nonumber \\
s(m)=m_is_i, \quad K(m)=m_iK_{ij}m_j,
\end{eqnarray}
where $\mathbf{m}$ is an integer valued vector, and repeated indices mean summation. We define further a set of local fields:
\begin{equation}
\label{e8}
\Psi_m=e^{-i\phi(m)},
\end{equation}
which obey
\begin{equation}
\label{e9}
\Psi_m(x)\Psi_{m'}(x')=(-1)^{q(m)q(m')}\Psi_{m'}(x')\Psi_m(x)
\end{equation}
for $x\neq x'$.
From the properties of $\mathbf{K}$ matrix in the symmetric representation we have:
\begin{equation}
\label{e10}
(-1)^{K(m)}=(-1)^{q(m)}
\end{equation}
Thus, if $K(m)$ is odd, the field $\Psi_m$ is fermionic, and if $K(m)$ is even, it is bosonic. Now we consider the tunneling perturbation:
\begin{equation}
\label{e11}
T=\int dx[t(x)\Psi_m(x)+h.c.].
\end{equation}
It should be bosonic and charge conserving, hence $q(m)=s(m)=0$ and $K(m)$ even. The scaling dimension of $\Psi_m$ is $\Delta(m)$ that satisfies the inequality
\begin{equation}
\label{e12}
\Delta(m)\geqslant\frac{1}{2}|K(m)|.
\end{equation}
If the tunneling perturbation is relevant, the modes in $\Psi_m$ become massive and are removed from the low energy theory. From the scaling perspective it is potentially relevant if the scaling dimension $\Delta(m)<2$. So, with the constraints given above, we conclude that the condition for $\mathbf{m}$ that leads to a potentially mass generating perturbation is  
\begin{equation}
\label{e13}
K(m)=q(m)=s(m)=0.
\end{equation}
We see that although the space may be flat, the spin vector still plays a role in the properties of edge states. 

We now apply this analysis to the edge states at the boundary of $\nu=\frac{2}{3}$ spin-polarized and unpolarized regions. The $\{\mathbf{K},\mathbf{q},\mathbf{s}\}$ of the effective theory for this state, where two phases coexist, is:
\begin{equation}
\label{e14}
\mathbf{K}=\left(\begin{array}{cccc}1 & 2 & 0 & 0 \\2 & 1 & 0 & 0 \\0 & 0 & -1 & -2 \\0 & 0 & -2 & -1\end{array}\right), 
\mathbf{q}=\left(\begin{array}{c}1 \\1 \\1 \\1\end{array}\right), 
\mathbf{s}=\left(\begin{array}{c}\frac{1}{2} \\\frac{1}{2} \\\frac{1}{2}\\-\frac{1}{2}\end{array}\right).
\end{equation}
In terms of CF theory, we can identify fields $\phi_i$ as $\phi_1$, $\phi_2$ corresponding to $\Lambda_{0\uparrow}$,  $\Lambda_{0\downarrow}$ respectively, and $\phi_3$, $\phi_4$ corresponding to $\Lambda_{0\uparrow}$,  $\Lambda_{1\uparrow}$ respectively. From Eqs. (\ref{e13}) and (\ref{e14}), we find out two independent solutions for $\mathbf{m}$: $\mathbf{m}_1=(1,0,-1,0)$ and $\mathbf{m}_2=(0,1,-1,0)$. The solution $\mathbf{m}_2$ represents tunneling between $\Lambda_{0\downarrow}$ and $\Lambda_{0\uparrow}$, which is unlikely to happen since there is an energy gap. Therefore, only the operator $\Psi_{m_1}$ is relevant and potentially mass generating, and $\phi_1$, $\phi_3$ are removed from the low energy theory.  Thus, we have shown that in the K-matrix description there are only two counter-propagating edge states with opposite spins. 

Quite remarkably on the level of K-matrix description, in the low energy sector of the domain wall, neutral modes do not emerge, and spin-charge separation of the spin-unpolarized phase does not appear. There are always questions of possible edge reconstruction, the role of disorder, and in this particular case, a problem of how the domain wall edges couple to the true edges at the boundary of the sample, which is even more complicated than behaviour of edge states in homogeneous phases in the corners of a sample. However, our conclusion on the two counter-propagating states with opposite spins in the domain wall, supported by the K-matrix description as well as by a handwaving crossed electric and effective magnetic fields argument, rings true. We shall see that this picture is also suppproted by the exact diagonalization in disk and torus geometry.

\section{III. Numerical calculations in the disk geometry}
Here we will use exact diagonalization in the disk geometry in order to confirm the conclusions of the previous section about the induced edge states on the boundary of the $\nu=\frac{2}{3}$ spin-polarized and spin-unpolarized regions.  Some of the results in this section  have already been briefly discussed  in our paper Ref.\cite{wu2017parafermion}. For completeness of our analysis, we include these here, with an improved numerical procedure and extended discussion.

\begin{figure}
\centering
\subfigure
{
\begin{minipage}[b]{0.22\textwidth}
\includegraphics[width=4.5cm,height=4cm]{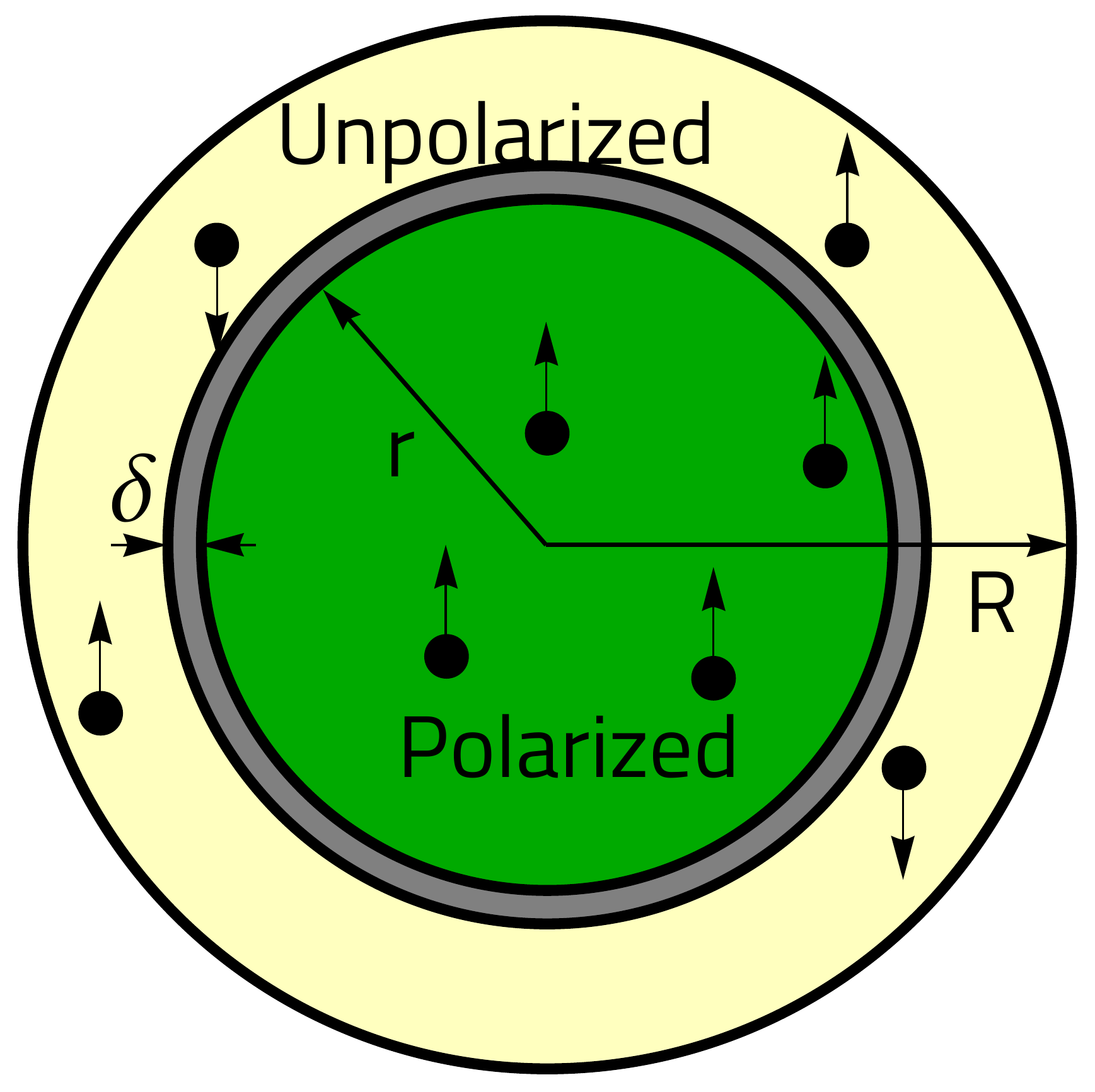}
\put(-130,110){(a)}
\end{minipage}

\label{3a}
}
\hfill
\subfigure
{
\begin{minipage}[b]{0.22\textwidth}
\includegraphics[width=4cm,height=4cm]{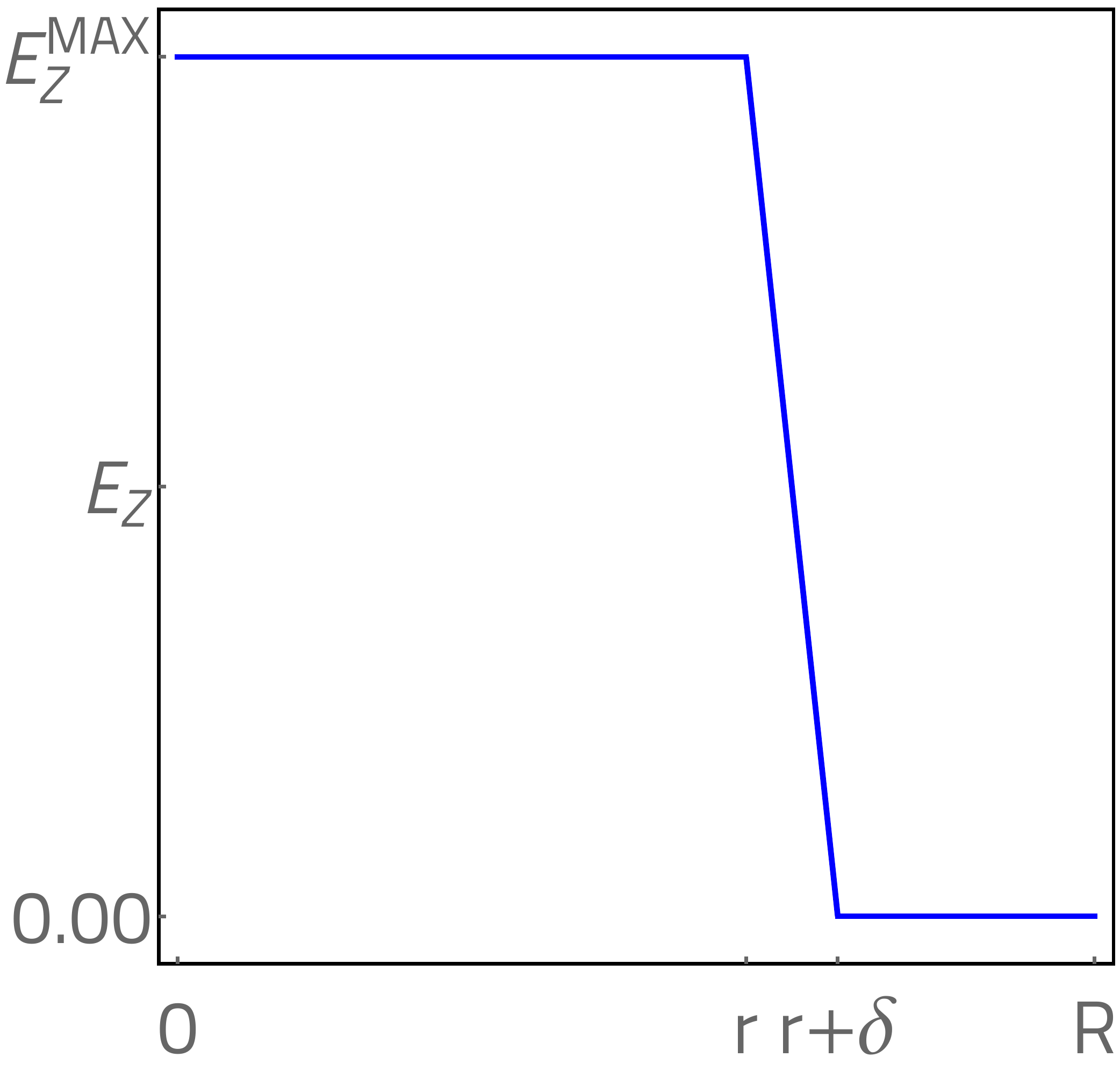}
\put(-120,110){(b)}
\end{minipage}
\label{3b}
}
\\
\subfigure
{
\begin{minipage}[b]{0.45\textwidth}
\includegraphics[width=7.0cm,height=4.5cm]{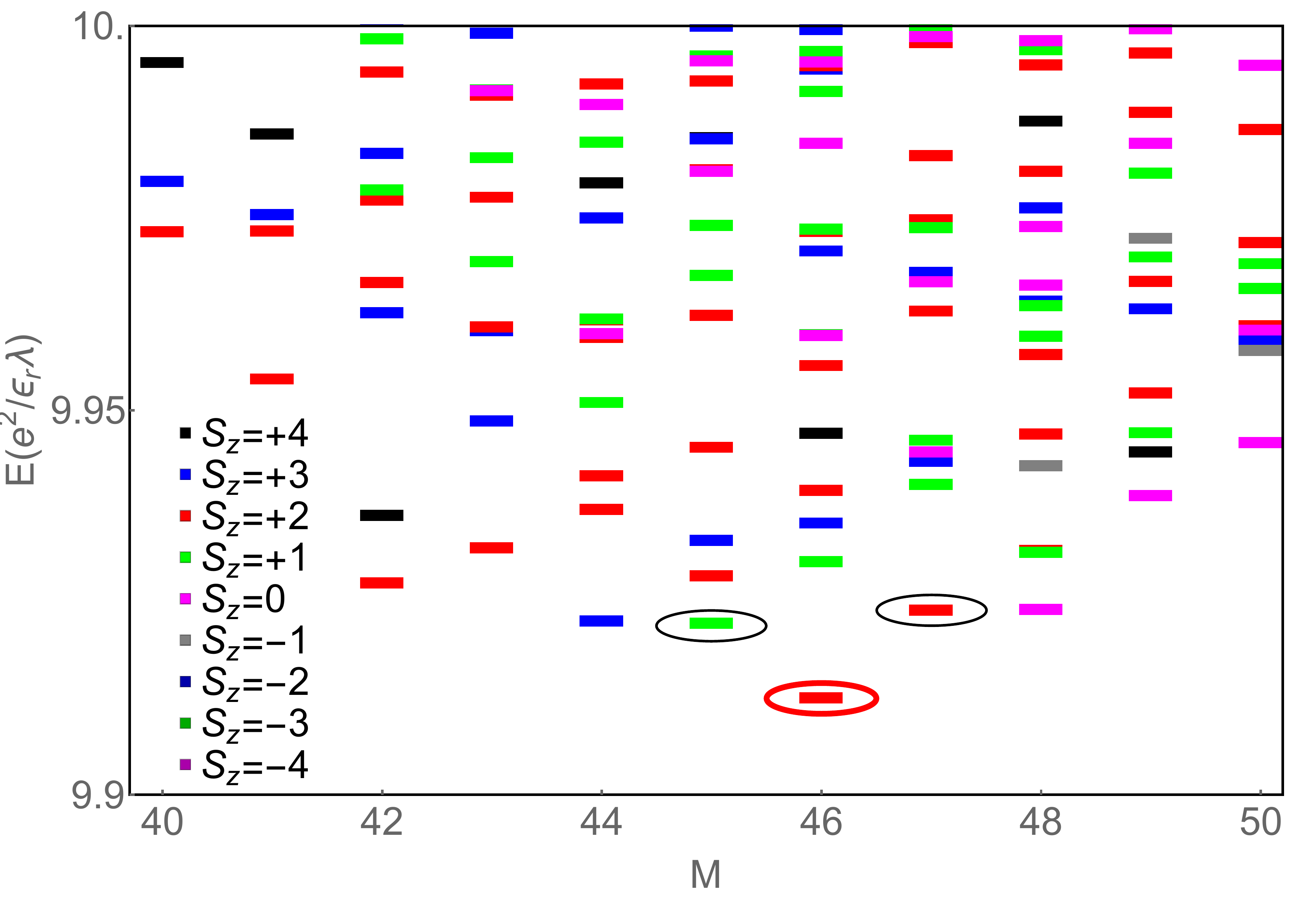}
\put(-232,110){(c)}
\end{minipage}
\label{3c}
}
\caption{(Color online). a: Disk geometry for the simulation domain. b: The profile of the Zeeman splitting of electron states. c: Spectra of 8 electrons on the disk with the profile of the Zeeman splitting shown in Fig.\ref{3b}. They are characterized  by an total angular momentum $L_z$ and a total electron spin $S_z$. The ground state with $L_z=46$ is circled red. Edge excitations with the same $S_z=2$ as in the ground state and with $L_z=45,47$, corresponding to the addition or subtraction of a single flux, are circled black. \label{fig3}}
\end{figure}

We simulate the system of 8 electrons in a magnetic field using the disk geometry shown schematically in Fig.\ref{3a}. In this model we use a spatially dependent Zeeman energy to control the spin polarization in $z-direction$, see Fig.\ref{3b}. The central region of the disk with a radius $R_1$ is characterized by a large Zeeman term $E_Z^{max}$, while the outer region with the outer diameter $R_2$ is set to $E_Z^{min}=0$. The Zeeman term is varied smoothly within the region defined by an interval of  $r$ given by $R_1<r<R_1+\Delta R$, resulting in a smooth variation of the wavefunctions across the disk and avoiding spurious effects originating from an abrupt change of the Zeeman splitting. When there are 8 electrons on the disk, the allowed single particle states have angular momentum $0\leqslant m \leqslant11$. Thus, $R_2=\sqrt{22}l_m=4.8l_m$. We design the central region in such a way that it contains half of the electrons, corresponding to the condition $R_1+\Delta R=\sqrt{11}l_m=3.3l_m$. We set $R_1=2.9l_m$ and $\Delta R=0.4l_m$. We anticipate that the resulting spin density will reflect that electrons should be spin-polarized in the central region and spin-unpolarized in the outer region in , e.g., the difference of spin densities in inner and outer regions. Note that due to a strong penetration of electron wavefunction from the outer $R_1<r<R_2$ region into the inner $r<R_1$ region of the disk, the central region will contribute significantly to average spin polarization of all states. We shall see that this contribution of the central region leads to a decrease in the difference of the average spin splitting $\int \psi(r)^*E_Z(r)\psi(r)d^2r$ for the modes on the two sides of the domain wall to about $<6\%$, similar to the experimental conditions of Ref.~\cite{wu2017parafermion}. 

The electron Hamiltonian is given by:
\begin{eqnarray}
\label{e17}
\mathcal{H}_d&= &\frac{1}{2m^*}\sum_i\left(\mathbf{p}+\frac{e\mathbf{A}}{c}\right)_i^2+E_z(r_i)\sigma_z^{(i)}+U_i\nonumber\\
&+&\sum_{ij}\frac{e^2}{\epsilon|r_i-r_j|}
\end{eqnarray}

\begin{figure}
\centering
\subfigure
{
\begin{minipage}[b]{0.22\textwidth}
\includegraphics[width=4cm,height=4cm]{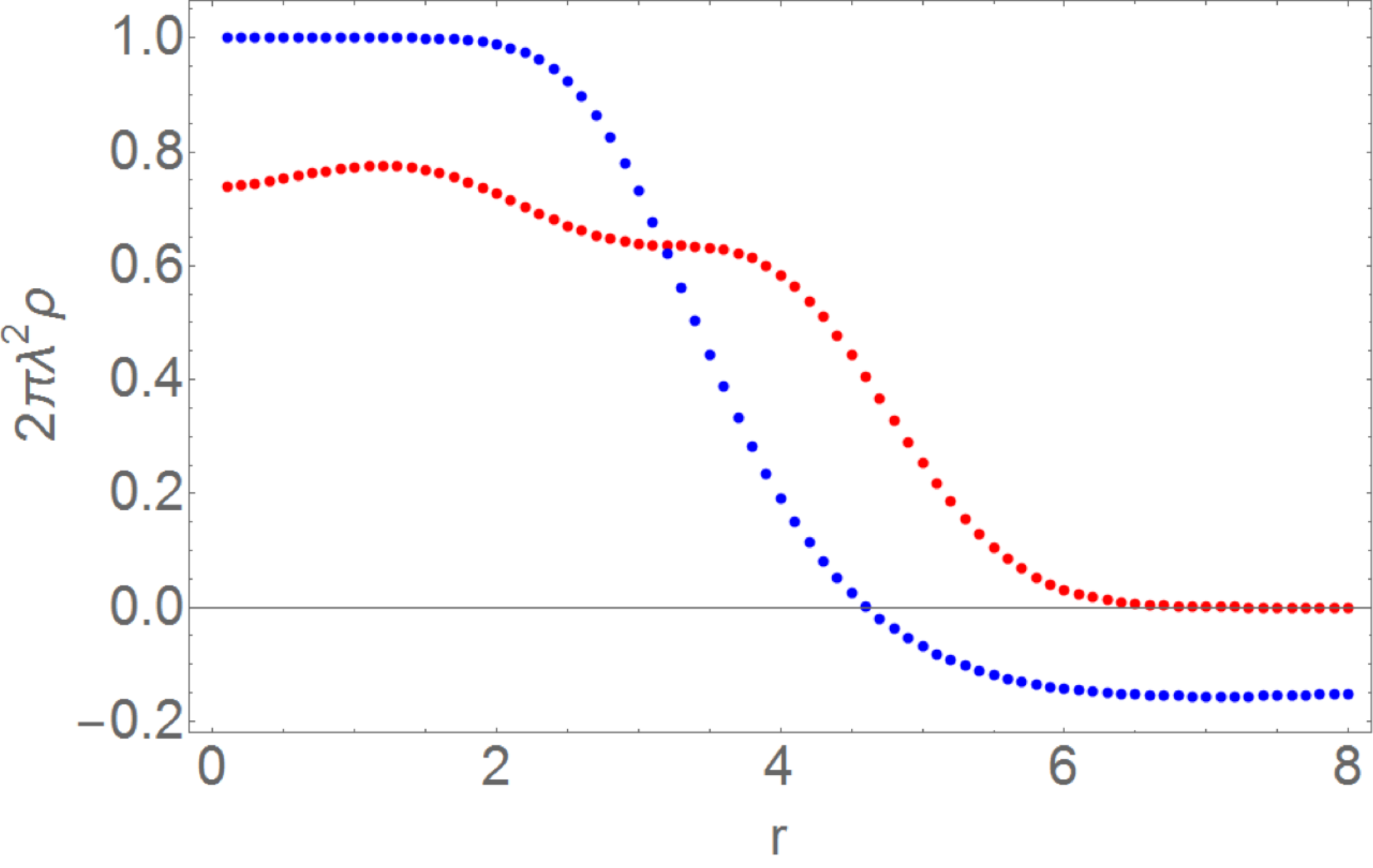}
\put(-130,110){(a)}
\end{minipage}

\label{4a}
}
\hfill
\subfigure
{
\begin{minipage}[b]{0.22\textwidth}
\includegraphics[width=4cm,height=4cm]{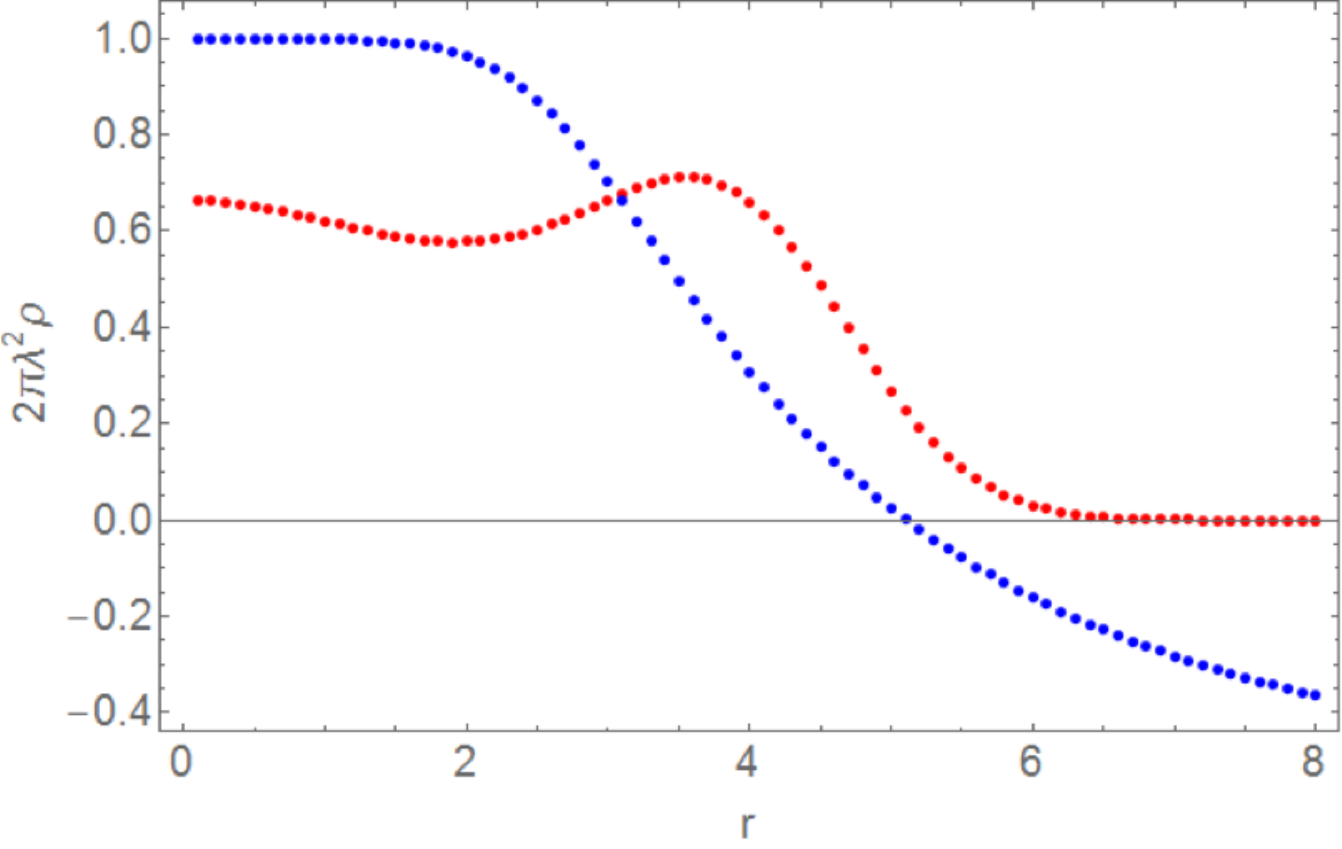}
\put(-120,110){(b)}
\end{minipage}
\label{4b}
}
\subfigure
{
\begin{minipage}[b]{0.22\textwidth}
\includegraphics[width=4cm,height=4cm]{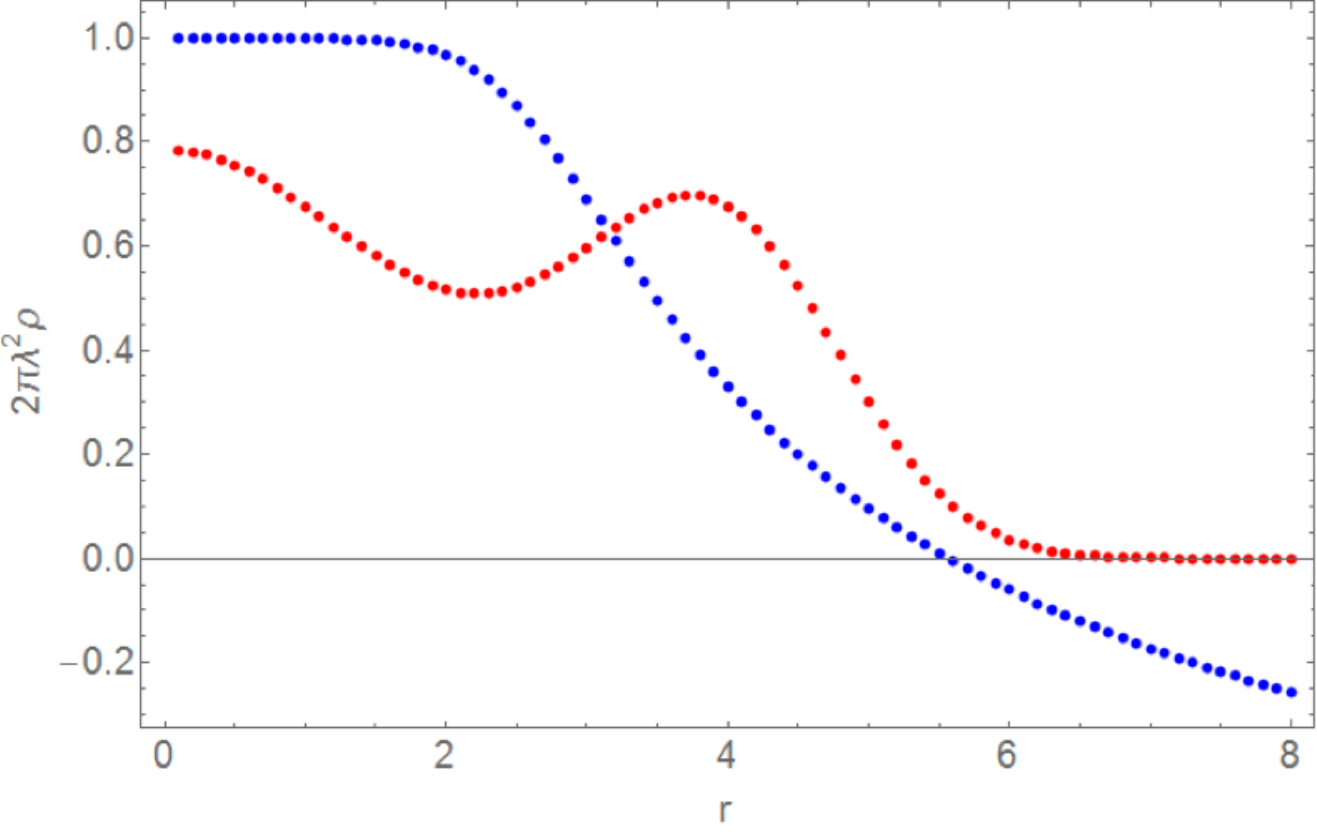}
\put(-130,110){(c)}
\end{minipage}

\label{4c}
}
\hfill
\subfigure
{
\begin{minipage}[b]{0.22\textwidth}
\includegraphics[width=4cm,height=4cm]{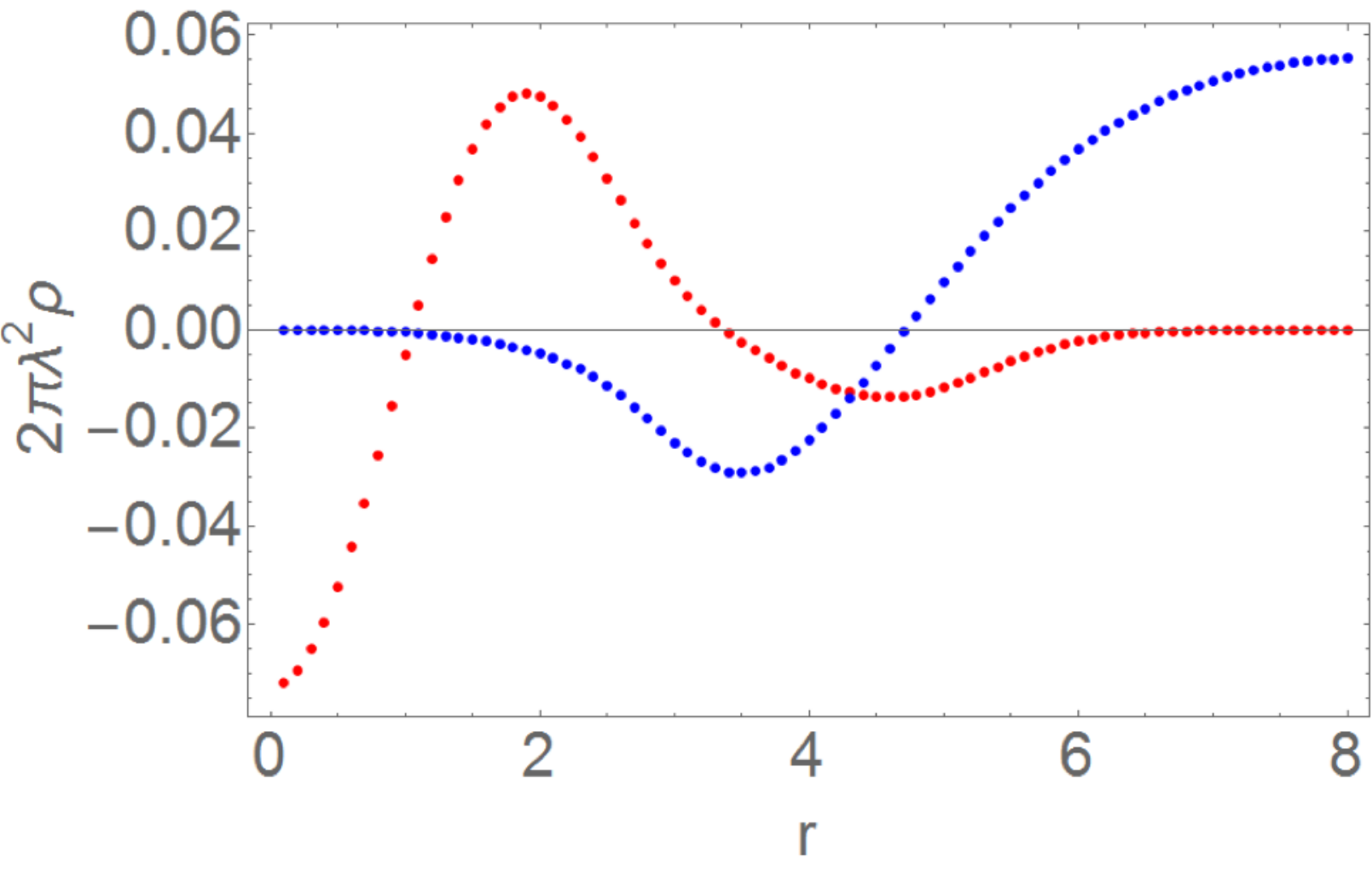}
\put(-120,110){(d)}
\end{minipage}
\label{4d}
}
\subfigure
{
\begin{minipage}[b]{0.22\textwidth}
\includegraphics[width=4cm,height=4cm]{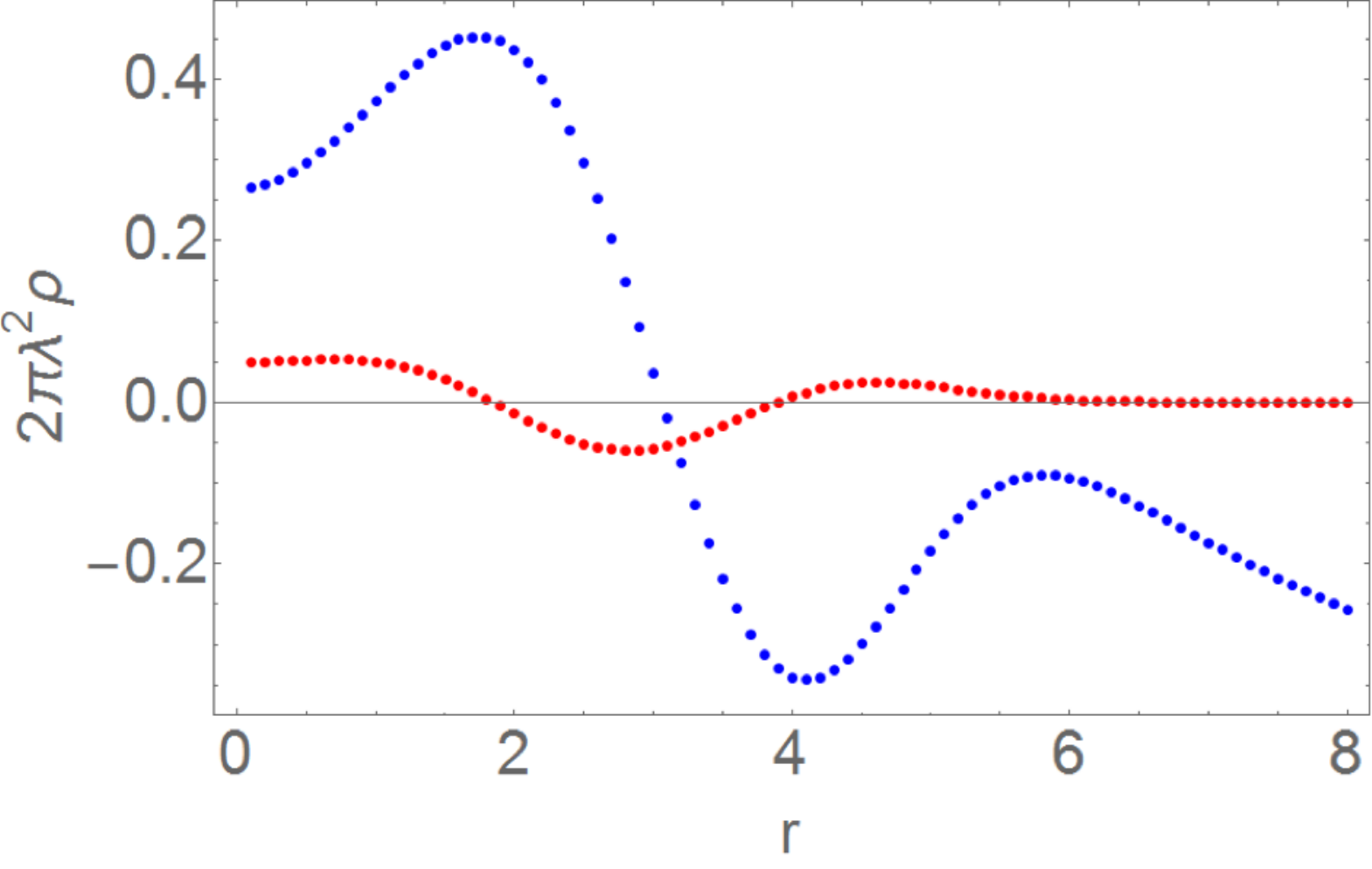}
\put(-130,110){(e)}
\end{minipage}

\label{4e}
}
\hfill
\subfigure
{
\begin{minipage}[b]{0.22\textwidth}
\includegraphics[width=4cm,height=4cm]{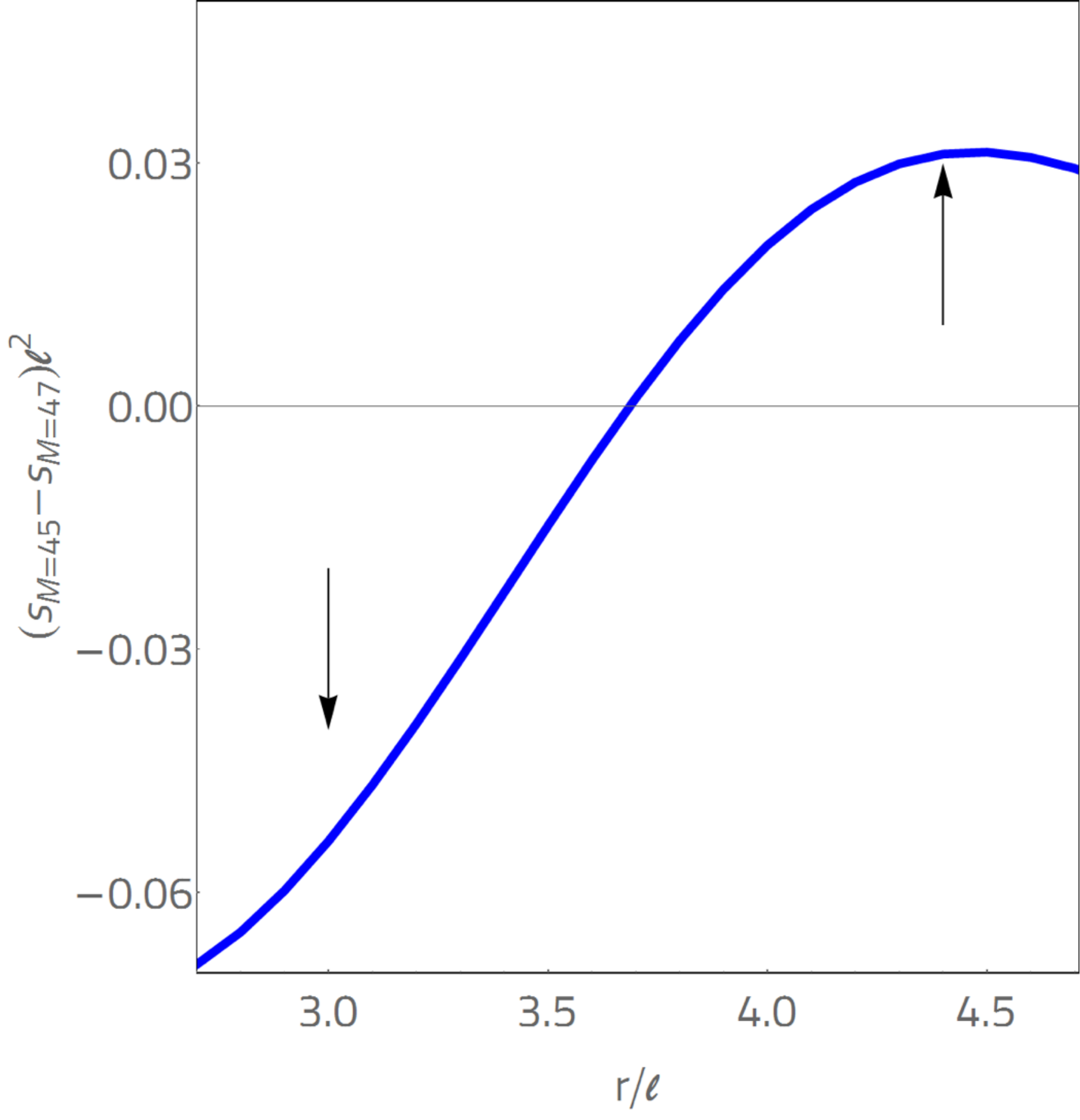}
\put(-120,110){(f)}
\end{minipage}
\label{4f}
}

\caption{(Color online). a: The ground state electron density (red) and spin density (blue) for 8 electrons on the disk containing the domain wall between polarized and unpolarized states at a filling factor 2/3 in a magnetic field. b: The density profile (red) and spin polarization (blue) for the edge state $M=45$. c: The density profile (red) and spin polarization (blue) for the edge state $M=47$. d: The differences of density (red) and spin (blue) between $M=45$ edge state and the ground state. e: The differences of density (red) and spin density (blue) between $M=47$ edge state and the ground state. f: Spin density difference between edge state $M=45$ and $M=47$. \label{fig4}}
\end{figure}

The first term and the second term in Eq.(\ref{e17}) are kinetic energy and Zeeman energy respectively. The third term is the parabolic confinement $U(r)=Cr^2$, confining electrons to the disk with $C=0.036e^2/\epsilon l_m^3$, and the last term is the Coulomb interaction between electrons. This Hamiltonian is diagonalized using a configuration interaction method. The states are classified by their projections of the total angular momentum on the z-axis, $L_z$, and the total spin of electrons, $S_z$. We exactly diagonalize this Hamiltonian for 8 electrons in a spatially varying Zeeman energy that models the coexistence of spin polarized and unpolarized states at a filling factor $\nu=2/3$. The exact diagonlization spectra are given in Fig.\ref{3c}. We have identified the ground state, which is spin-polarized in the center and unpolarized in the outer region of the disk, as well as the edge states flowing close to the boundary between spin polarized and spin unpolarized regions. Their number and spin density distributions are calculated. All of the results are shown in Fig.\ref{4a}-\ref{4f}.

The ground state has a total angular momentum $L_z=46$ and a total spin $S_z=2$. The total spin indicates that 6 electrons are in spin up states and 2 electrons are in spin down states, as expected. In the CF picture, there are N electrons with N/2 occupying $\Lambda_{0\uparrow}$ and N/4 occupying $\Lambda_{1\uparrow}$ in the center region and $\Lambda_{0\downarrow}$ in the outer region. The total angular momentum of the ground state is:
\begin{eqnarray}
\label{e18}
L_z&=&pN(N-1)+L_z^{CF}=N(N-1)\nonumber\\
&-&(\frac{N}{4}(\frac{N}{2}-1)+(\frac{N}{4}-3)\frac{N}{8}+(\frac{3N}{4}-1)\frac{N}{8}\nonumber\\
&=&N(N-1)-\frac{N(N-3)}{4}=\frac{N(3N-1)}{4}
\end{eqnarray}
For N=8, $L_z=46$ indeed, coinciding with our numerical result. The ground state is separated by a gap from the rest of the spectra as shown Fig.\ref{3c}, and does not carry the electric current. From the spin profile in Fig.\ref{4a}, the ground state is indeed spin-polarized in the center of the disk and spin-unpolarized in the outer region. 

The lowest energy excitations that have spin polarization of the ground state and correspond to a substraction or addition of a single flux have $L_z=45$ and $L_z=47$, see Fig.\ref{4b} and Fig.\ref{4c}. These are the modes that carry an electrical current. When compared to the ground state, they have $\Delta L=-1$ and $\Delta L=1$ respectively. This indicates that these two edge states have opposite components of linear velocities. The differences in density and in the spin polarization density between the two edge states and the ground state are shown in Fig.\ref{4d} and Fig.\ref{4e}. We observe that the density differences are large only around the internal edge, which confirms that these two edge states correspond to the internal boundary between regions with large and small Zeemann interactions, i.e. the domain wall. In Fig.\ref{4f}, we show the results for the difference of spin densities of the two modes near the domain wall between polarized and unpolarized region. Despite the finite size effects in a small system, the exact diagonalization clearly identifies that the two edge states in the domain wall area have components of spin density with opposite orientation. Our numerical study clearly shows that there are two counter-propagating edge states in the domain wall with different spin polarizations, which is consistent with our analysis in Sec.II.

\section{IV. Numerical calculations on the torus}
In this section, we will numerically study the system in a torus geometry. The advantage of the torus geometry is that it allows to avoid considering the edge between the fractional Quantum Hall liquid and a vacuum that is present in the disk configuration. Hence  physics of the induced edge between spin polarized and unpolarized regions is elucidated. 

The torus geometry is represented as a rectangular cell with periodic boundary conditions. This geometry has been considered in Ref.\cite{yoshioka1983ground} for the $\frac{1}{3}$ FQH state. We apply the method \cite{yoshioka1983ground} to our case. We take the coordinate system such that the boundary of the rectangular cell is given by $x=0, x=a, y=0, y=a$, with the vector potential $\overrightarrow{\mathbf{A}}=(0, xB)$. We have $2\pi l_B^2\frac{N}{a^2}=\nu$, therefore $a=\sqrt{24\pi}l_B=8.68l_B$, and there are $m=\frac{N}{\nu}=12$ single electron orbitals in the cell. 
\begin{figure}
\centering
\subfigure
{
\begin{minipage}[b]{0.22\textwidth}
\includegraphics[width=4cm,height=4cm]{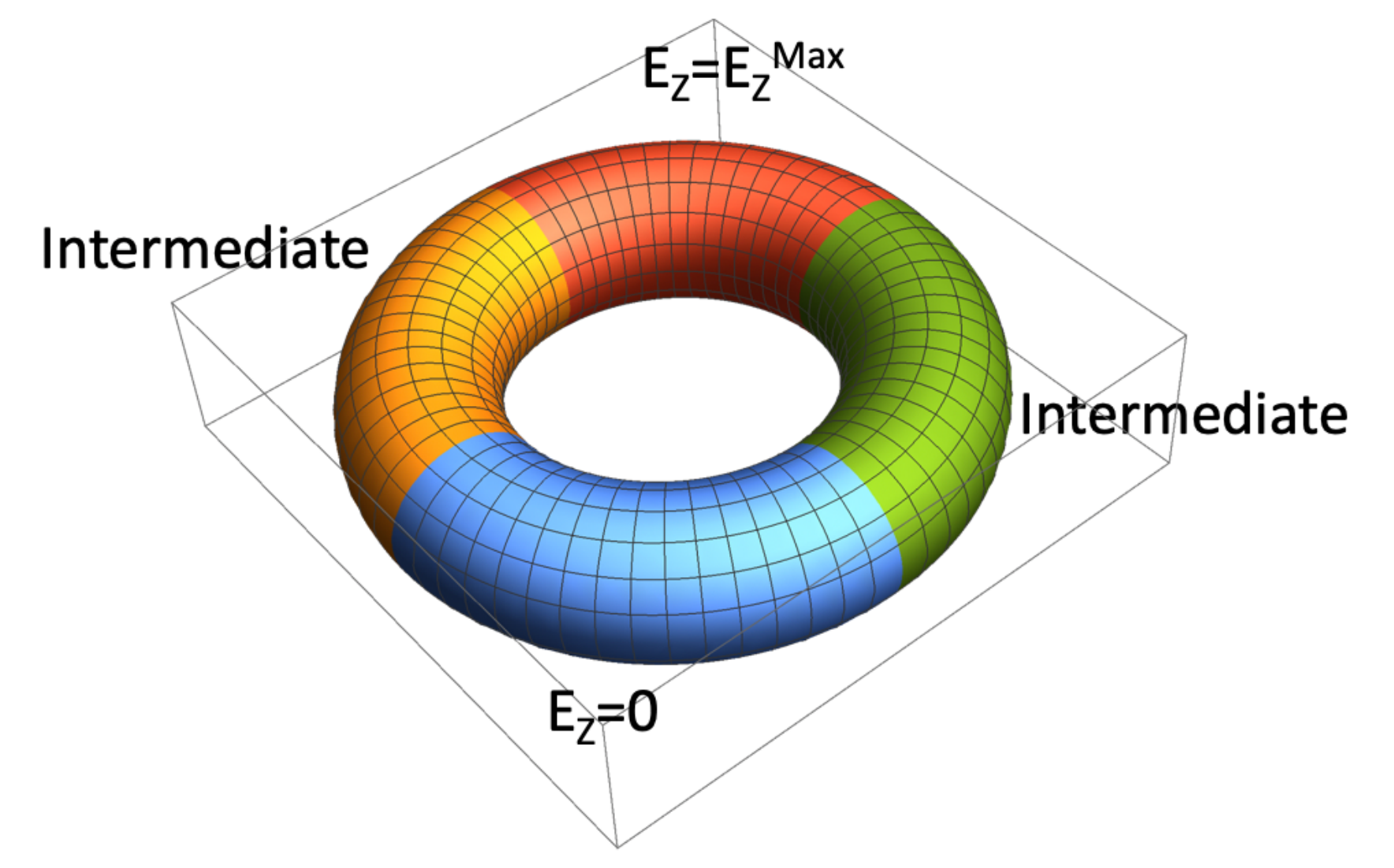}
\put(-130,110){(a)}
\end{minipage}

\label{5a}
}
\subfigure
{
\begin{minipage}[b]{0.22\textwidth}
\includegraphics[width=4cm,height=4cm]{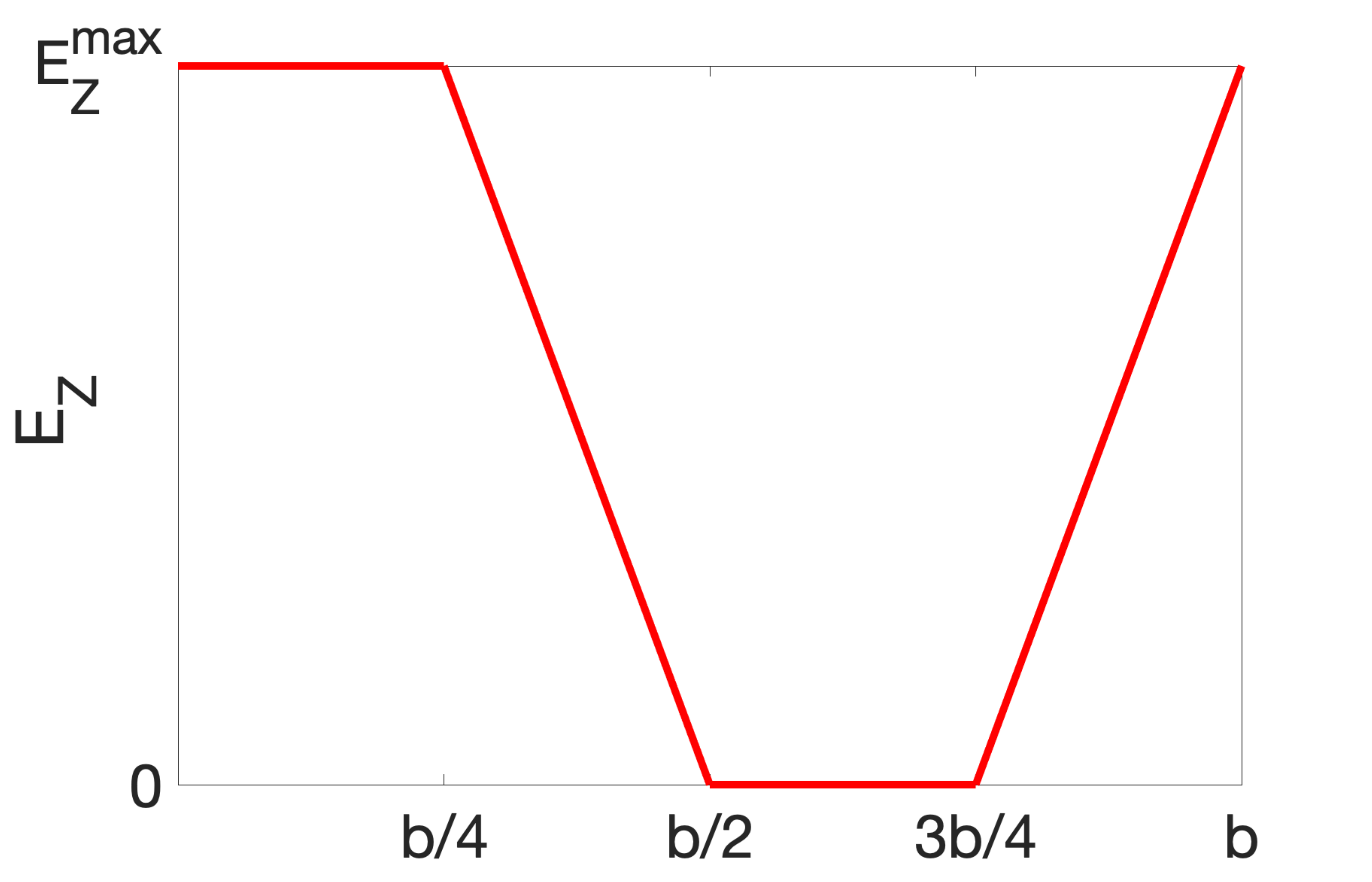}
\put(-120,110){(b)}
\end{minipage}
\label{5b}
}
\subfigure
{
\begin{minipage}[b]{0.45\textwidth}
\includegraphics[width=8cm,height=5cm]{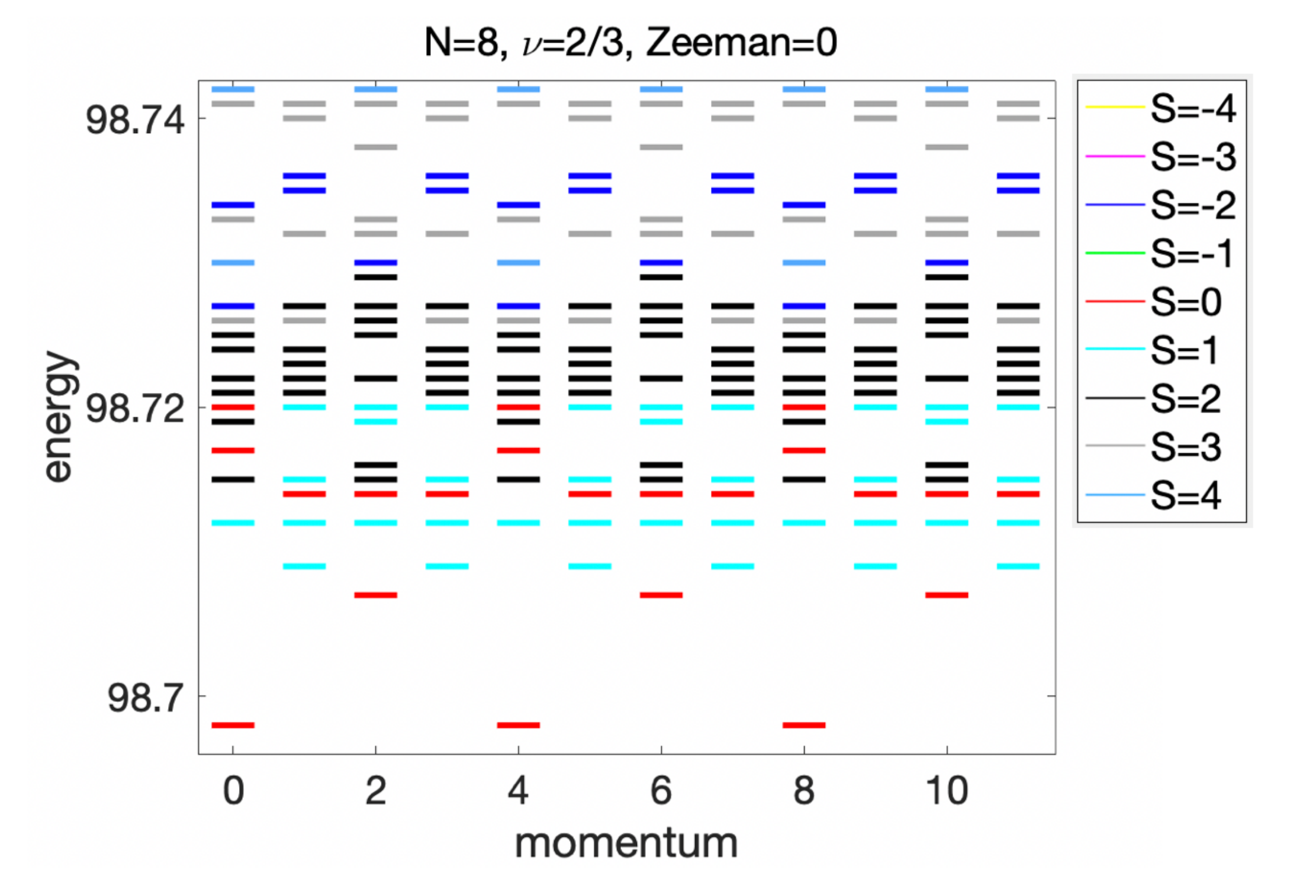}
\put(-232,110){(c)}
\end{minipage}

\label{5c}
}
\\
\subfigure
{
\begin{minipage}[b]{0.45\textwidth}
\includegraphics[width=8cm,height=5cm]{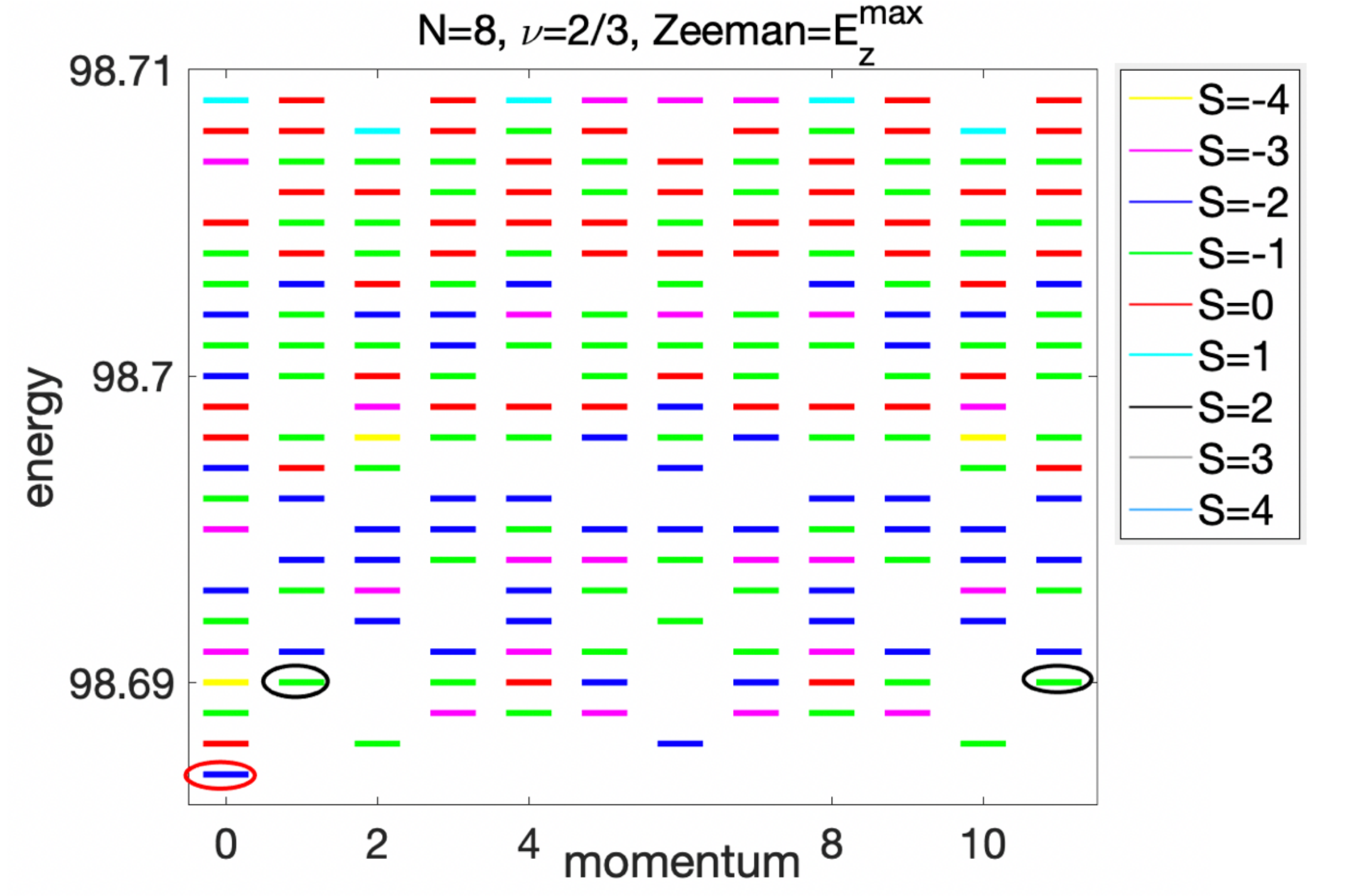}
\put(-232,110){(d)}
\end{minipage}
\label{5d}
}

\caption{(Color online). a: The torus geometry for a spatially varying Zeeman energy. b: The amplitude of Zeeman energy along the toroidal direction. c: Spectra of 8 electrons on the torus without Zeeman splitting. The ground state has three-fold degeneracy. d: Spectra of 8 electrons on the torus with profile of Zeeman energy shown in Fig.\ref{5b}. Electrons are characterized  by a total angular momentum (mod 12) $L_z$ and a total spin $S_z$ of particles. Ground state is the $L_z=0$ and $S_z=2$ state, circled red. Edge excitations with the same $S_z=2$ as in the ground state with $L_z=1,11$, which correspond to the addition or subtraction of a single flux, are circled black.  \label{fig5}}
\end{figure}
The wavefunctions of these orbitals are given by:
\begin{equation}
\label{e19}
\phi_j(\overrightarrow{r})=(\frac{1}{a\pi^{1/2}l_B})^{\frac{1}{2}}\sum_{k=-\infty}^{\infty}e^{[i\frac{(X_j+ka)y}{l_B^2}-\frac{(X_j+ka-x)^2}{2l_B^2}]}
\end{equation}
where $j$ labels the $j-$th orbit, $1\leqslant j\leqslant m$, and $X_j=\frac{j}{m}a$ is the coordinate of the guiding center. The Hamiltonian of the system can be written as: 
\begin{eqnarray}
\label{e20}
\mathcal{H}_t&= &\frac{1}{2m^*}\sum_i\left(\mathbf{p}+\frac{e\mathbf{A}}{c}\right)_i^2+E_z(r_i)\sigma_z^{(i)}\nonumber\\&+&\sum_{ij}V(\mathbf{r}_i-\mathbf{r}_j).
\end{eqnarray}
The first and the second terms are the kinetic and Zeeman term, correspondingly. The third term is the Coulomb interaction of electrons, and due to the boundary conditions, it is given by:
\begin{eqnarray}
\label{e21}
V(\mathbf{r})=\sum_s\sum_t \frac{e^2}{\epsilon|\mathbf{r}+sa\hat{x}+ta\hat{y}|}.
\end{eqnarray}
The Coulomb matrix elements are determined by:
\begin{eqnarray}
\label{e22}
&V_{j_1j_2j_3j_4}=\nonumber\\
&\frac{1}{2}\int d^2r_1d^2r_2
\phi_{j_1}^*(\mathbf{r}_1)\phi_{j_2}^*(\mathbf{r}_2)V(\mathbf{r}_1-\mathbf{r}_2)\phi_{j_3}(\mathbf{r}_3)\phi_{j_4}(\mathbf{r}_4)\nonumber\\
&=\frac{1}{2a^2}\frac{2\pi e^2}{\epsilon q}\sum_q'\sum_s\sum_t \delta_{q_x,\frac{2\pi s}{a}}\delta_{q_y,\frac{2\pi t}{a}}\delta_{j_1-j_4,t}'\times \nonumber\\
&\exp[-\frac{l_B^2q^2}{2}-2\pi is\frac{j_1-j_3}{m}]\delta_{j_1+j_2, j_3+j_4}'.
\end{eqnarray}
Here the symbols with prime are defined modulo $m$ and the summation over $q$ excludes $q=0$. From the above expression we observe that the total angular momentum is conserved only modulo $m$. Therefore, we are going to use the total angular momentum $M(mod m) $ and the total spin $S$ to classify the quantum states. 

We first exactly diagonalize the 8-electron Hamiltonian in the lowest Landau level in the absence of the Zeeman term. In this case, only the Coulomb terms play a role. The spectra are shown in Fig.\ref{5c}. We find that the ground state state has degeneracy three, which is consistent with Ref.\cite{wen1990ground}, in which the degeneracy is shown to be given by $|det(K)|$ . Eq.(\ref{e4}) indeed gives the degeneracy three. 

Now we turn on the spatially dependent Zeeman term defined by Fig.\ref{5a} and \ref{5b}. We divide the torus into four equal regions. One of these regions has large Zeeman energy $E_Z^{max}$, while the opposite side of the torus  is subject to zero Zeeman energy. Zeeman energy varies smoothly in the regions between these two  from zero to $E_Z^{max}$. Exact diagonalization leads to spectra shown in Fig.\ref{5d}. We observe that there is a single ground state with $M=0$ and $S=2$ circled red in Fig.\ref{5d}. The reason for the lifted degeneracy is a broken symmetry of magnetic translations  in the presence of the spatially varying Zeeman term.

\begin{figure}
\centering
\subfigure
{
\begin{minipage}[b]{0.22\textwidth}
\includegraphics[width=4cm,height=4cm]{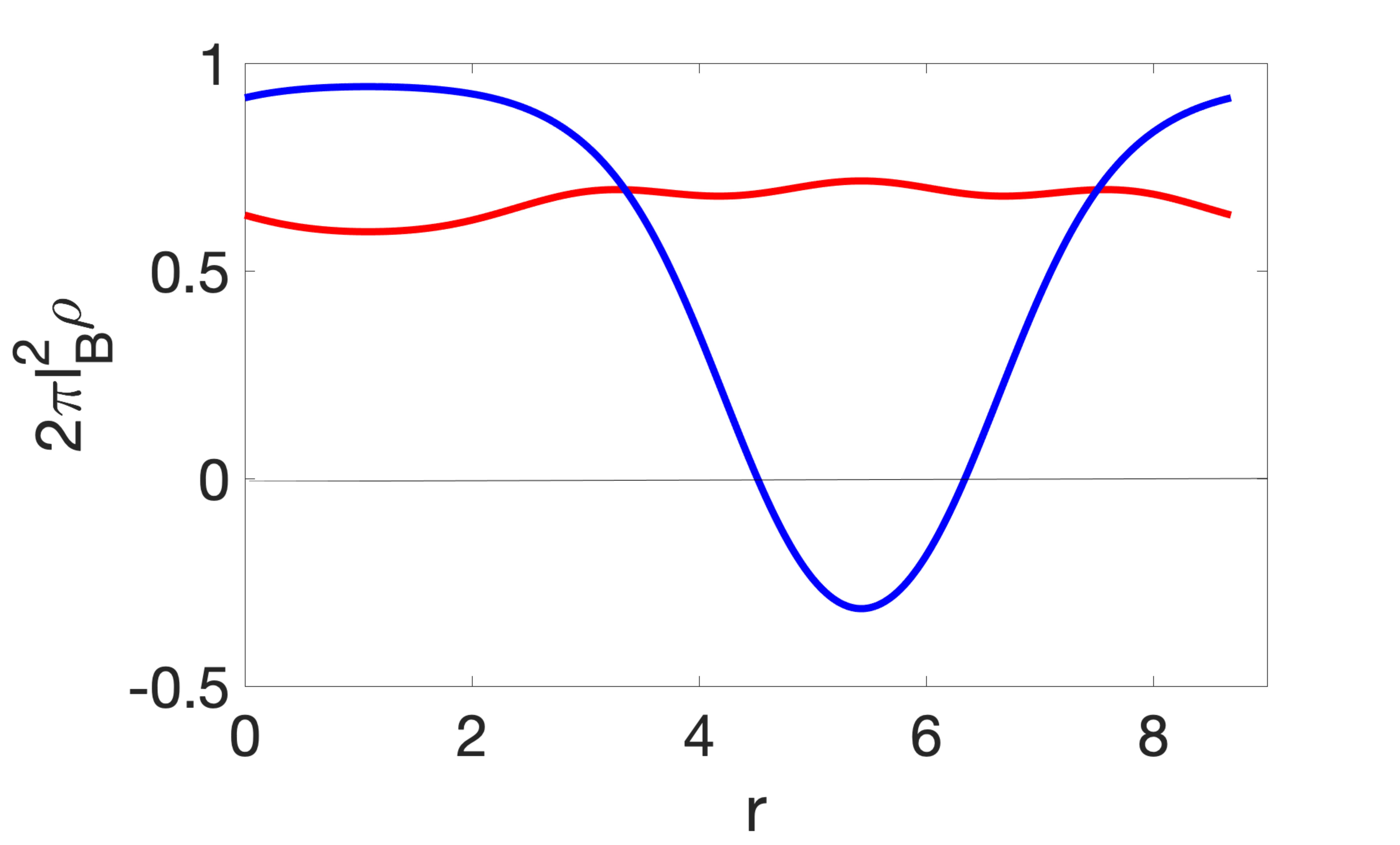}
\put(-130,110){(a)}
\end{minipage}

\label{6a}
}
\hfill
\subfigure
{
\begin{minipage}[b]{0.22\textwidth}
\includegraphics[width=4cm,height=4cm]{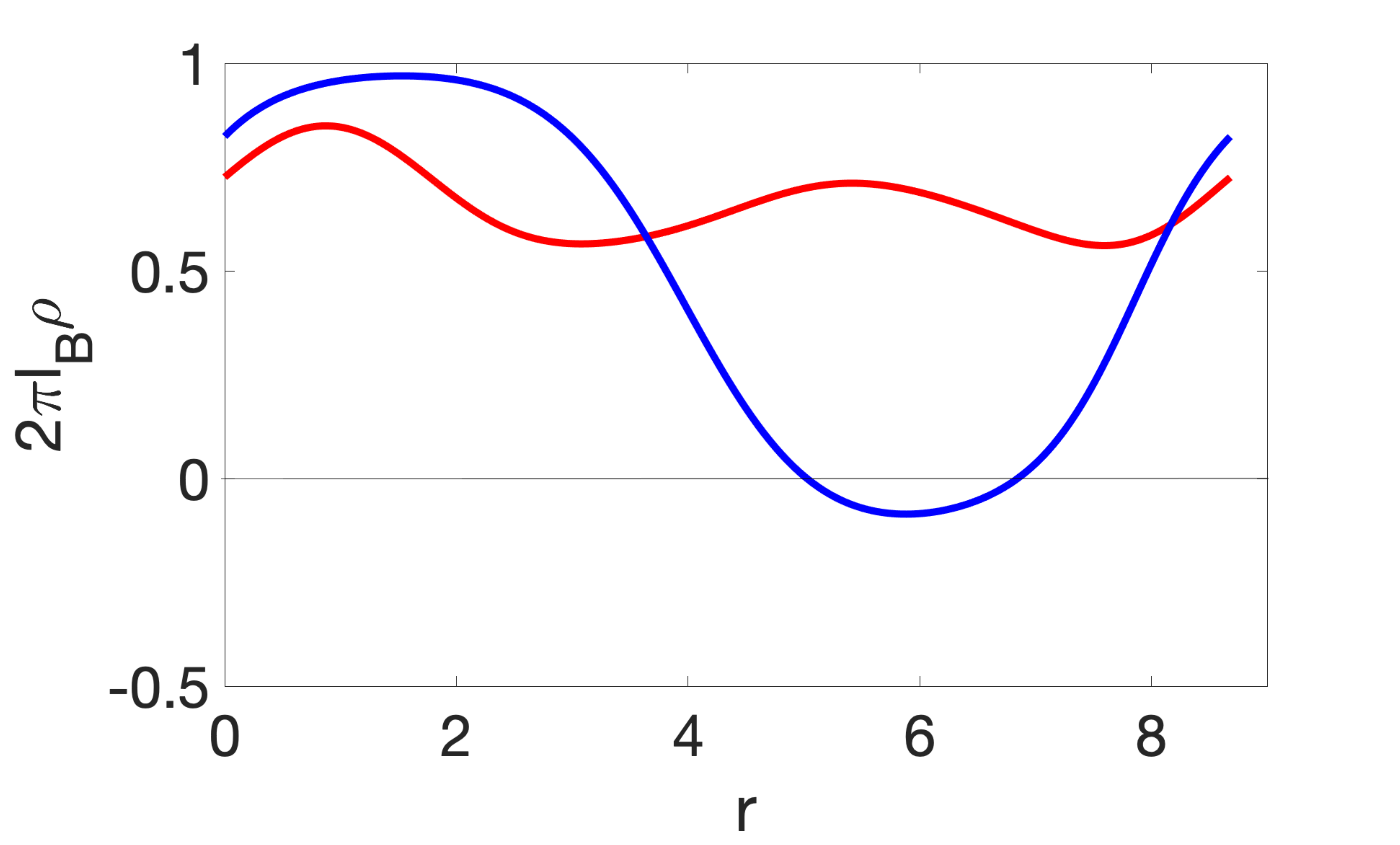}
\put(-120,110){(b)}
\end{minipage}
\label{6b}
}
\subfigure
{
\begin{minipage}[b]{0.22\textwidth}
\includegraphics[width=4cm,height=4cm]{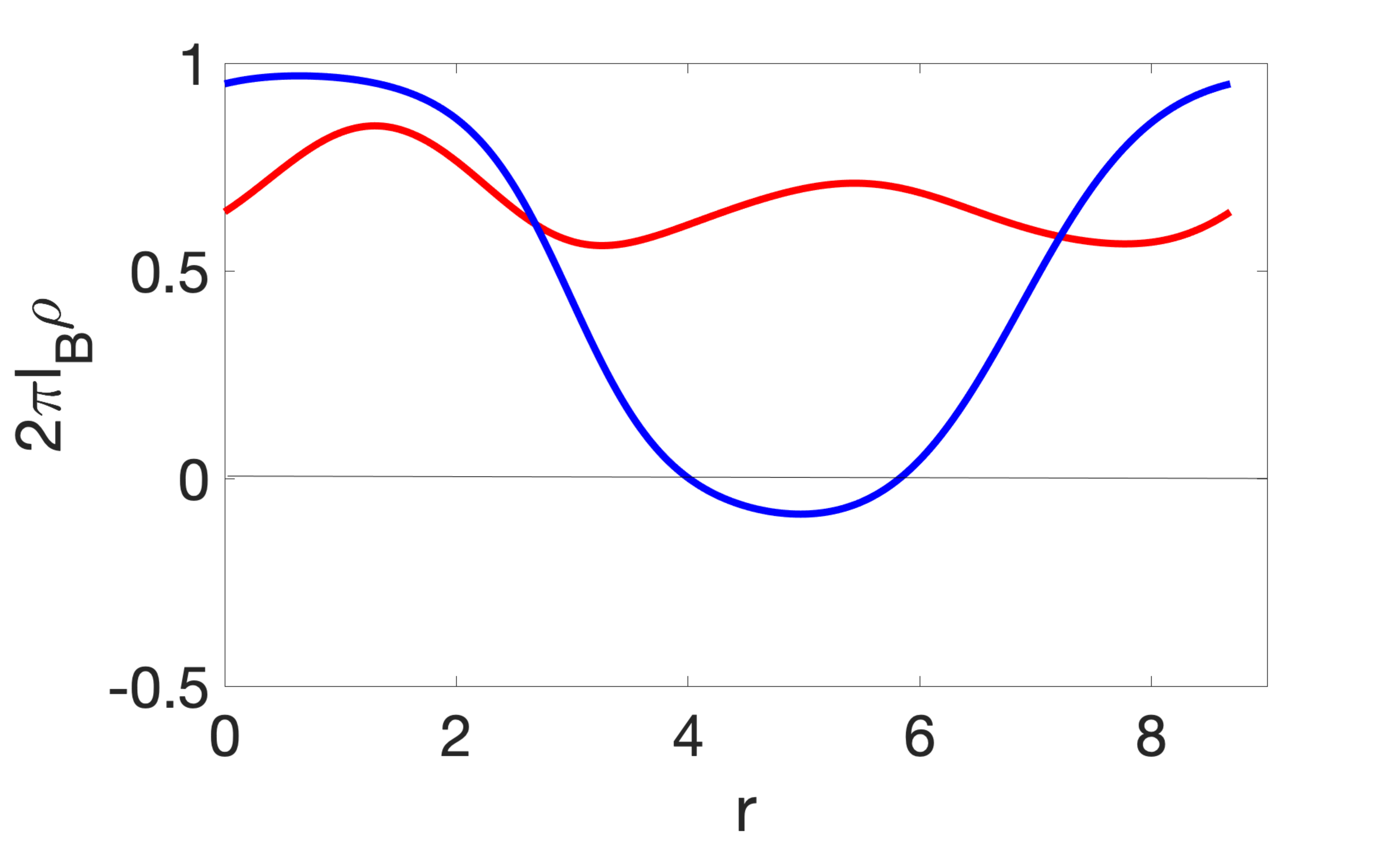}
\put(-130,110){(c)}
\end{minipage}

\label{6c}
}
\hfill
\subfigure
{
\begin{minipage}[b]{0.22\textwidth}
\includegraphics[width=4cm,height=4cm]{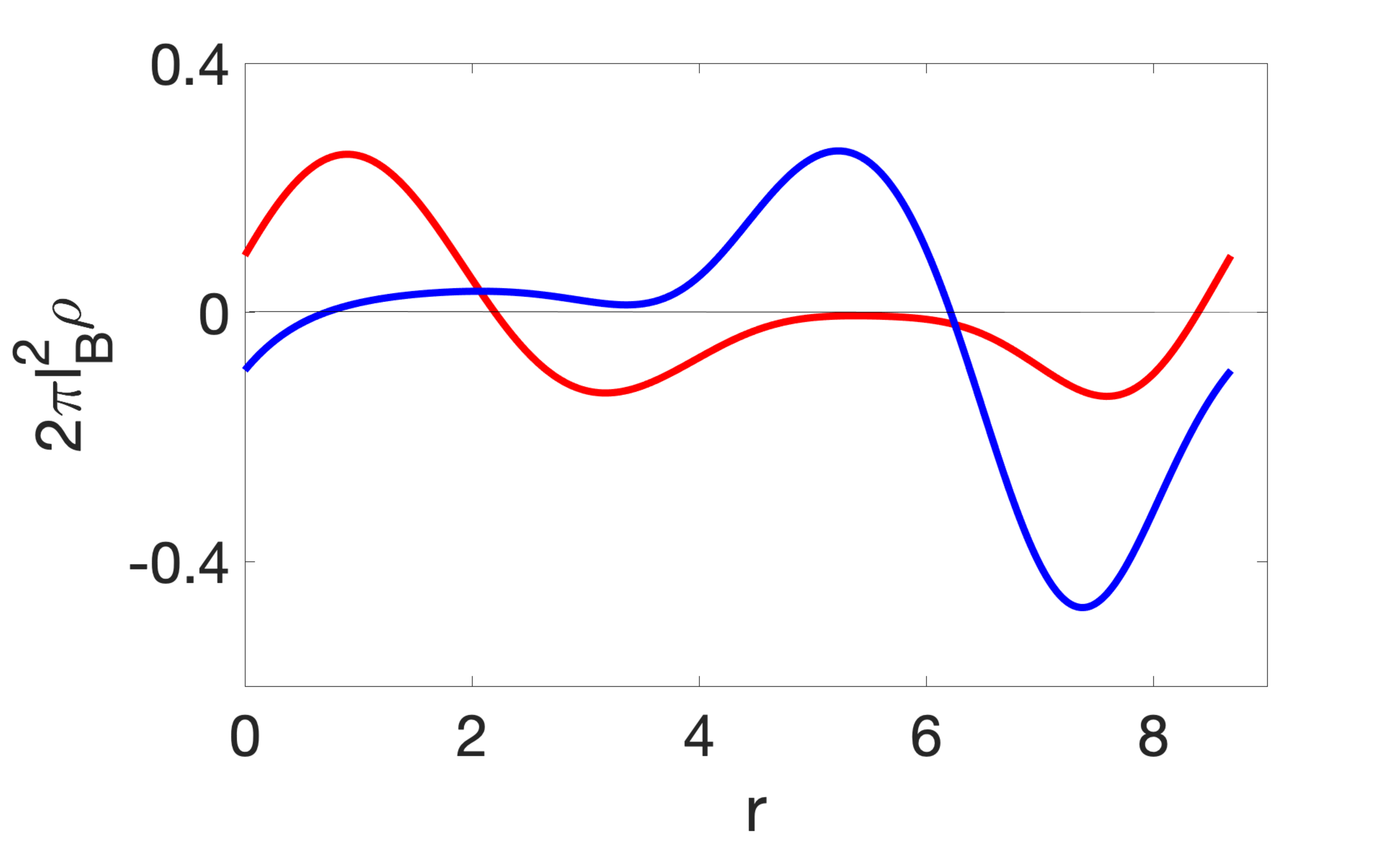}
\put(-120,110){(d)}
\end{minipage}
\label{6d}
}
\subfigure
{
\begin{minipage}[b]{0.22\textwidth}
\includegraphics[width=4cm,height=4cm]{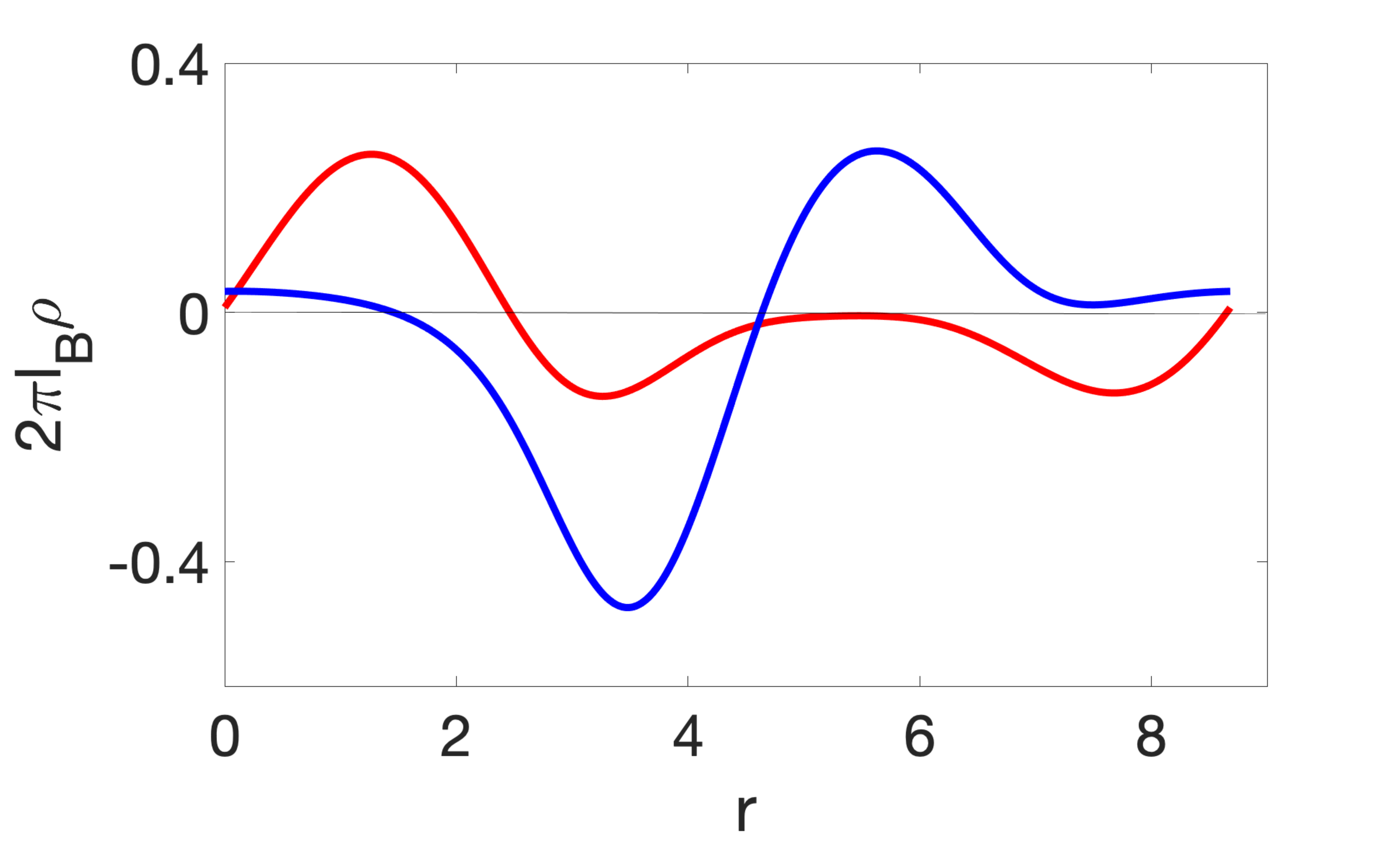}
\put(-130,110){(e)}
\end{minipage}

\label{6e}
}
\hfill
\subfigure
{
\begin{minipage}[b]{0.22\textwidth}
\includegraphics[width=4cm,height=4cm]{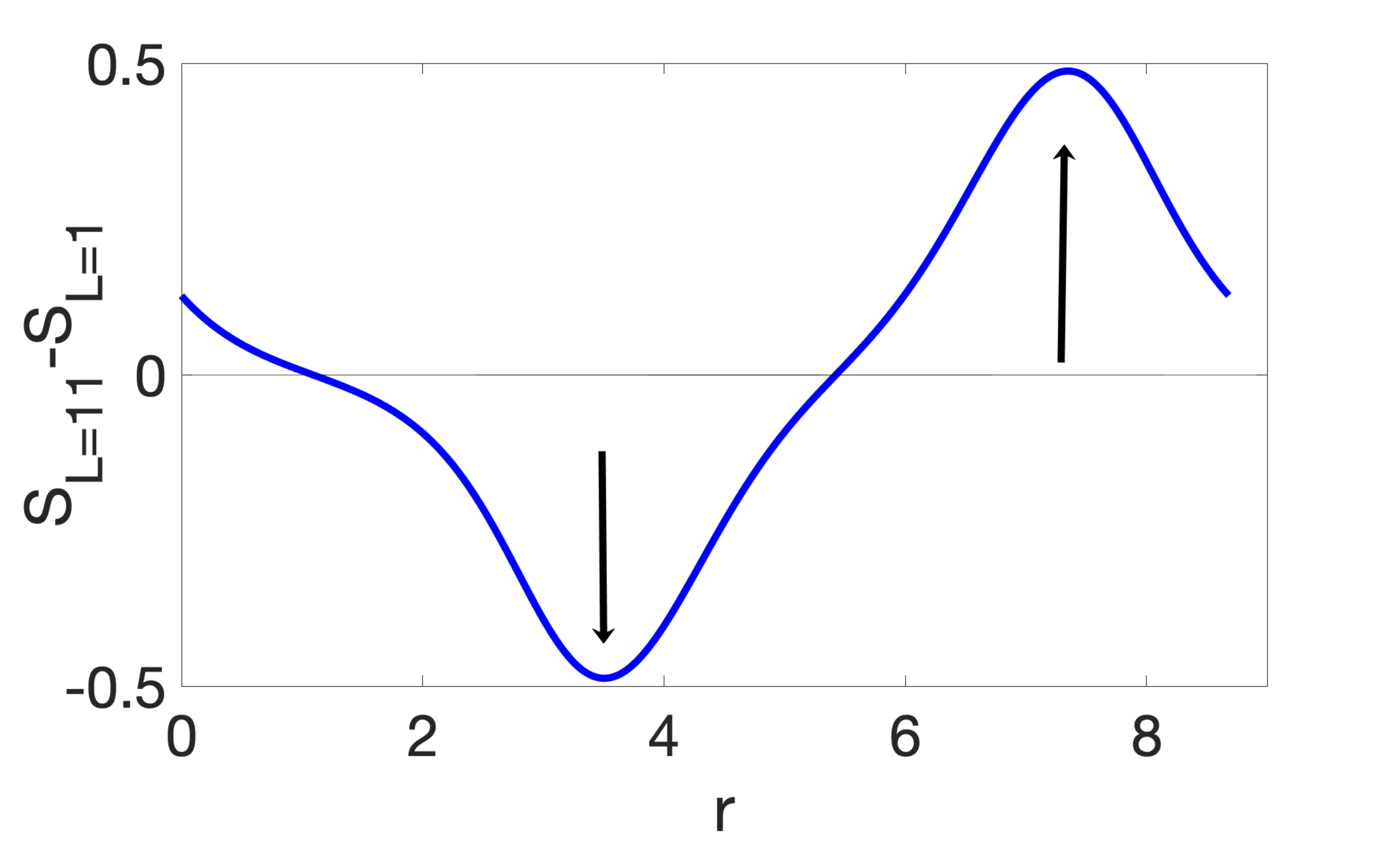}
\put(-120,110){(f)}
\end{minipage}
\label{6f}
}

\caption{(Color online). a: The ground state electron density (red) and spin density (blue) for 8 electrons on a torus containing the domain wall between spin-polarized and unpolarized states at a filling factor 2/3 in a magnetic field. b: The density profile (red) and spin polarization (blue) for the edge state $M=1$. c: The density profile (red) and spin polarization (blue) for the edge state $M=11$. d: The differences of density (red) and spin (blue) between $M=1$ edge state and the ground state. e: The differences of density (red) and spin (blue) between $M=11$ edge state and the ground state. f: Spin difference between edge state $M=11$ and $M=1$. \label{fig6}}
\end{figure}
We find the density profile and the spin polarization of the ground state, shown in Fig.\ref{6a}. The density is fluctuating slightly around $\nu=2/3$, as expected. The spin polarization is almost unity within the region where $E_Z=E_Z^{max}$ (region A) and has a dip in the region $E_Z=0$ (region B). This clearly indicates that the electrons are spin-polarized in the region A and 
spin-unpolarized in the region B. Therefore, our numerical calculation indeed simulate the state in which spin-polarized and 
spin-unpolarized fractional quantum Hall phases coexist. 

We now study the edge states. Comparison of Fig.\ref{5c} and Fig.\ref{5d} shows several low energy excitations. We are most interested in the two states with the same total spin as the total spin in the ground state. The two states correspond to to single edge state quanta flowing in the positive and negative poloidal directions. These states are circled black in Fig.\ref{5d}. Their total spin equals 2, and their angular momenta are $L=1$ and $L=11$, respectively. Their density distributions and spin polarizations, as well as the differences between these densities and those in the ground state are calculated, see Fig.\ref{6b} - Fig.\ref{6e}. In order to compare the spin polarizations of the edge states, we also calculate the difference in spin polarizations $S_{11}-S_1$, plotted in Fig.\ref{6f}. We see indeed that a domain wall (spin transition) emerges
between regions A and B. Therefore, our numerical study on the torus also supports the conclusion that there are two counter-propagating edge states with opposite spin polarizations in the domain wall between spin-polarized and spin-unpolarized states.

\section{V. Emergence of parafermion modes}
From the qualitative arguments, analytic theory and numerical calculations, we found that the edge states have opposite components of velocity and spin. Therefore, these states can potentially be coupled to an s-wave superconductor, a pre-requisite for generating topological superconductivity.  In the integer quantum Hall ferromagnets, proximity superconducting coupling has resulted in topological superconductivity in the domain wall region and in Majorana zero modes at the boundaries between topological and trivial superconducting regions\cite{simion2017disorder}. In the FQH regime, we anticipate the emergence of parafermions due to the fractional charges and fractional statistics of states comprizing the domain wall in much the same way as in \cite{clarke2013exotic}. In this section, we will quantitatively show how the parafermions emerge and can be controlled when coupled to an s-wave superconductor in the presence of spin-flipping interactions discussen in Appendix.

The physics of the edge modes is described by the action Eq.(\ref{e5}) with $K$ matrix given by Eq.(\ref{e14}). To simplify the expressions, we redefine the fields $\phi_{11}=\phi_1$, $\phi_{12}=\phi_2$, $\phi_{21}=\phi_3$, $\phi_{22}=\phi_4$. After quantizing these fields, we have the following commutation relations~\cite{wen2004quantum,mong2014universal}:
\begin{eqnarray}
\label{e23}
[\phi_{1\alpha}(x), \phi_{1\beta}(x')]=i\pi[(K^{-1})_{\alpha\beta}\mathrm{sgn}(x-x')+i\sigma_{\alpha\beta}^y],
\end{eqnarray}
\begin{eqnarray}
\label{e24}
[\phi_{2\alpha}(x), \phi_{2\beta}(x')]=i\pi[(-K^{-1})_{\alpha\beta}\mathrm{sgn}(x-x')+i\sigma_{\alpha\beta}^y],
\end{eqnarray}
\begin{eqnarray}
\label{e25}
[\phi_{1\alpha}(x), \phi_{2\beta}(x')]=i\pi[(-K^{-1})_{\alpha\beta}+i\sigma_{\alpha\beta}^y].
\end{eqnarray}
From the analysis of Sec.II, the remaining edge modes are generated by the fields $\phi_2$ and $\phi_4$. From Eq.(\ref{e23}) to (\ref{e25}), we find their commutation relations:
\begin{equation}
\label{e26}
[\phi_2(x),\phi_2(x')]=\frac{i\pi}{3}\mathrm{sgn}(x-x'),
\end{equation}
\begin{equation}
\label{e27}
[\phi_4(x),\phi_4(x')]=-\frac{i\pi}{3}\mathrm{sgn}(x-x'),
\end{equation}
\begin{equation}
\label{e28}
[\phi_4(x),\phi_2(x')]=\frac{i\pi}{3}.
\end{equation}
Therefore, $\phi_2$ and $\phi_4$ satisfy exactly the same commutation relations as $\phi_R$ and $\phi_L$ in Ref.\cite{clarke2013exotic}. We now discuss the emergence of parafermions. We observe that the path that lead to parafermions in \cite{clarke2013exotic} cannot work in the present case. The reason is, the spin-orbit interactions for electrons in the ground level are exceedingly small (see appendix) and cannot sufficiently gap the two counterpropagating modes. We will analize the possibility to generate transitions between counterpropagating edges with opposite spin by applying the in-plane magnetic field. It is important that the orbital part of the wavefunctions of the two counterpropagating are nearly the same,  a small difference exists only because these states are subject to opposite electric fields, due to the gradient of Zeeman splitting of the opposite spin states in the domain wall region. Thus, it is sufficient to mix spins by an in-plane magnetic field in order to generate transitions between the edge states. This hybridization results in the tunneling gap. Coupling the domain wall area to a conventional s-type superconductor, we introduce all ingredients for emergence of parafermions, and can apply a general argument discussed in  \cite{clarke2013exotic}. 

\begin{figure}
\centering
\subfigure
{
\begin{minipage}[b]{0.22\textwidth}
\includegraphics[width=4.5cm,height=4cm]{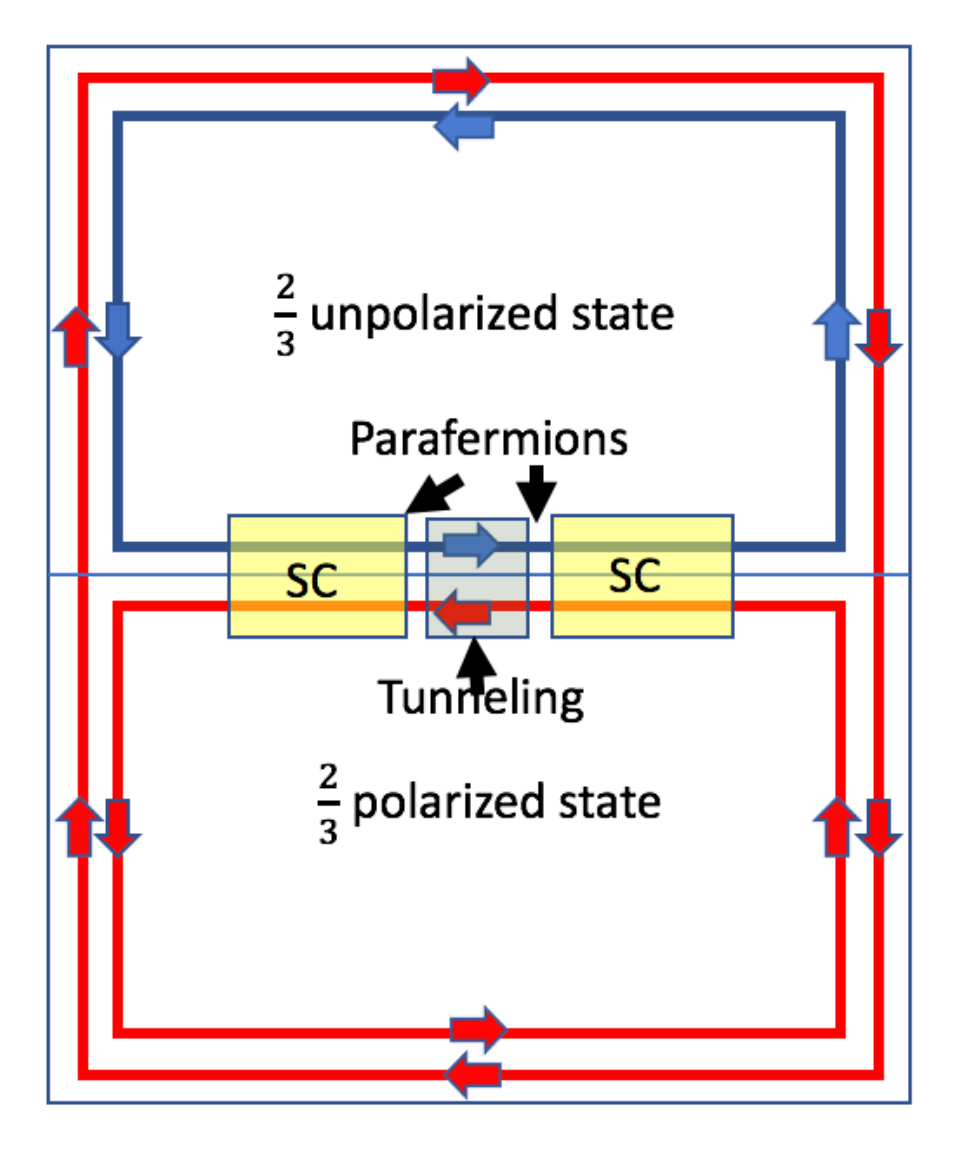}
\put(-130,110){(a)}
\end{minipage}

\label{7a}
}
\hfill
\subfigure
{
\begin{minipage}[b]{0.22\textwidth}
\includegraphics[width=3.8cm,height=4cm]{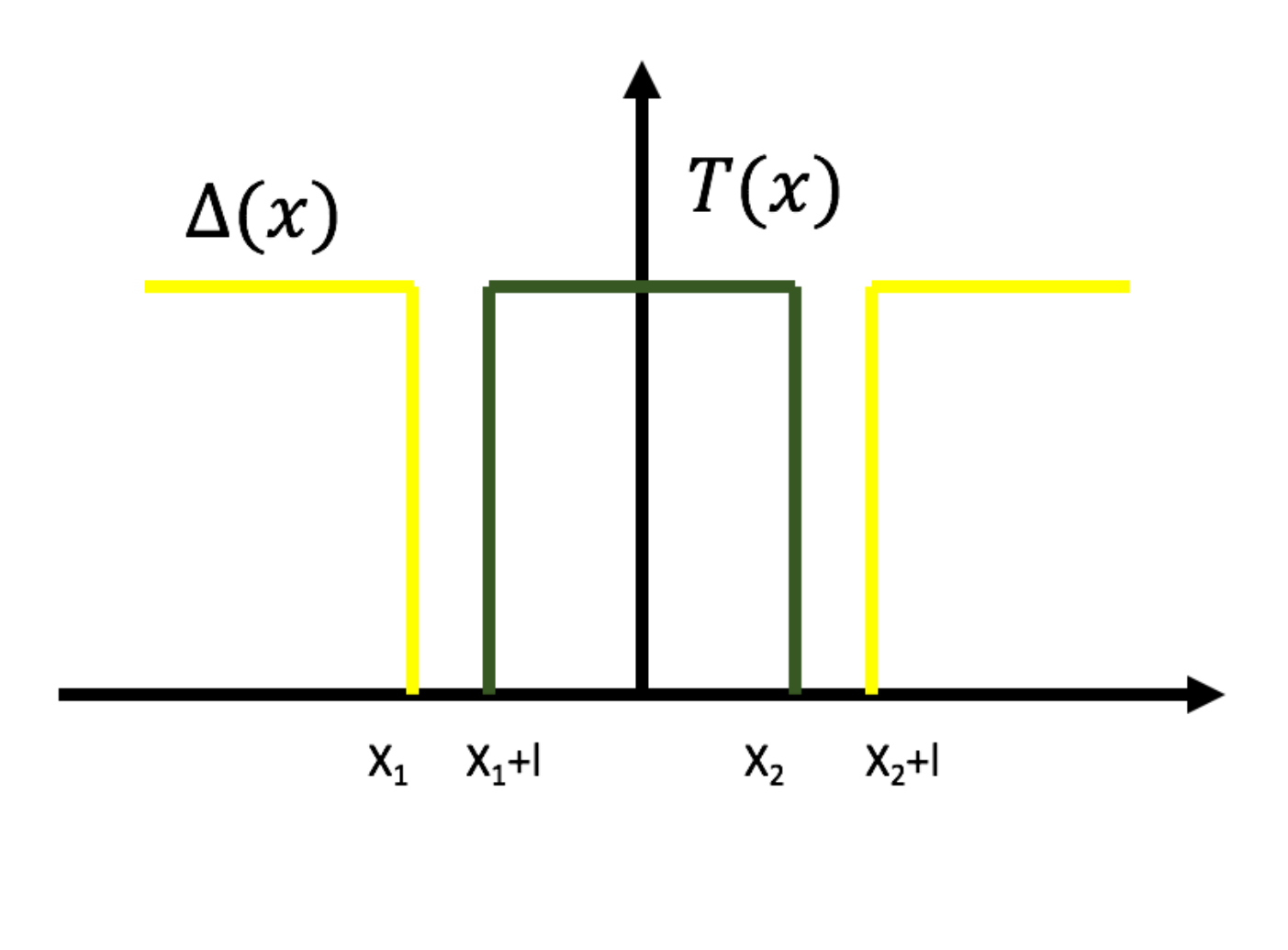}
\put(-115,110){(b)}
\end{minipage}
\label{7b}
}
\caption{(Color online). a: A schematic plot of experimental realization of parafermion zero modes. b: The spatial profile of the superconducting pairing and Zeeman-induced amplitudes $\Delta(x)$ and $T(x)$ induced by proximity effects in Fig.\ref{7a}. \label{fig7}}
\end{figure}

We envision the following architecture of the parafermion setting. For the proximity superconductivy, preliminary considerations show that proximity coupling to the edge quantum Hall states requires contacts with small Shottki barriers, allowing to control propagation of the Cooper pairs and electrons by the applied voltage. We assume that a proximity superconductors contact the domain wall area from the sides, as opposed to traditionally invisioned superconductor on the top of the semicoducting wire, quantum dot or a quantum well. Side configuration has additional advantage of gradually changing induced superconducting coupling from the contact to the region inside the domain wall area. When the competition of superconducting and tunneling gap leading to transition from trivial to topological proximity superconductivity depends on their relative value, spatial dependence of the superconducting coupling will allow to to tune the boundary between normal and topological superconductivity, where parafermions are expected to reside, by tuning this relative value. However, to simplify our consideration of emergence of parafermions, we follow \cite{clarke2013exotic} and consider the architecture in Fig.\ref{7a}, assuming for simplicity that two trivial superconducting regions are separated by a region, which produces a spin flip between edge states, i.e. the tunneling gap. The spatial profile of the pairing potential and tunneling for this simplified picture is given in Fig. \ref{7b}. We redefine the fields $\phi_{2/4}=\varphi\pm\theta$. The Hamiltonian of the interface is given by $H=H_0+H_1$, where 
\begin{equation}
\label{e29}
H_0=\frac{mv}{2\pi}\int dx[(\partial_x\varphi)^2+(\partial_x\theta)^2],
\end{equation}
$m=3$, and
\begin{equation}
\label{e30}
H_1\sim\int dx[-\Delta(x) cos(2m\varphi)-T(x)cos(2m\theta)].
\end{equation}
Assuming that angles $\theta$ and $\varphi$ obey $\varphi_{x<x_1}=\pi n_\varphi^1/m$, $\theta_{x\in(x_1+l,x_2)}=\pi n_\theta/m$, $\varphi_{x>x_2+l}=\pi n_\varphi^2/m$, we have:
\begin{equation}
\label{e31}
[n_\varphi^2, n_\theta]=i\frac{m}{\pi}.
\end{equation}
At low energy, we can focus on the interval between $x_j$ and $x_j+l$ governed by the effective Hamiltonian
\begin{equation}
\label{e32}
H_{eff}=\frac{mv}{2\pi}\sum_{i=1}^2\int_{x_i}^{x_i+l}dx[(\partial_x\varphi)^2+(\partial_x\theta)^2].
\end{equation}
We identify the operators
\begin{equation}
\label{e33}
a_j\rightarrow e^{i(\pi/m)(n_\varphi^j+n_\theta)},
\end{equation}
which commute with $H_{eff}$ and represent zero modes bound to areas between superconducting regions and the region where the gap between edge states is induced by tunneling. These modes obey the following relations:
\begin{equation}
\label{e34}
a_j^{2m}=1, a_ja_{j'}=a_{j'}a_je^{i(\pi/m)sgn(j'-j)}.
\end{equation}
Therefore, they are parafermion operators producing the $2m-$fold ground state degeneracy.

\section{VI. Numerical Calculations of  the Parafermion Zero Modes}
Having demonstrated that parafermions emerge in the simple model of the previous section, we now show numerically that parafermions arise when an s-wave superconductivity and tunneling are added to the quantum Hall states in a microscopic model. In the numerical simulation here, the appearance of parafermion modes is indicated by an emergence of six-fold degenerate ground state.

\begin{figure}
\centering
\subfigure
{
\begin{minipage}[b]{0.22\textwidth}
\includegraphics[width=4.5cm,height=4cm]{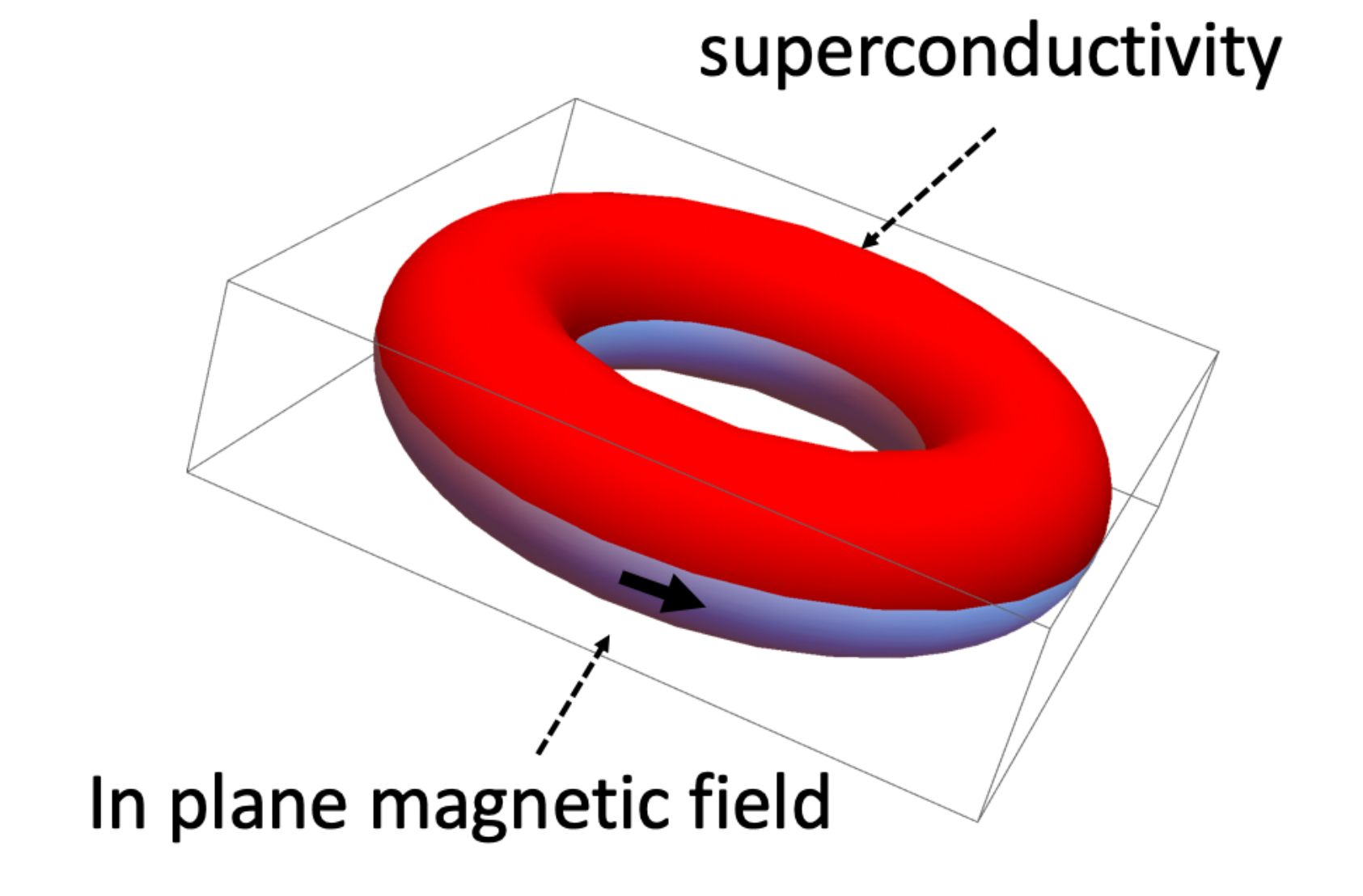}
\put(-130,110){(a)}
\end{minipage}

\label{8a}
}
\hfill
\subfigure
{
\begin{minipage}[b]{0.22\textwidth}
\includegraphics[width=3.8cm,height=4cm]{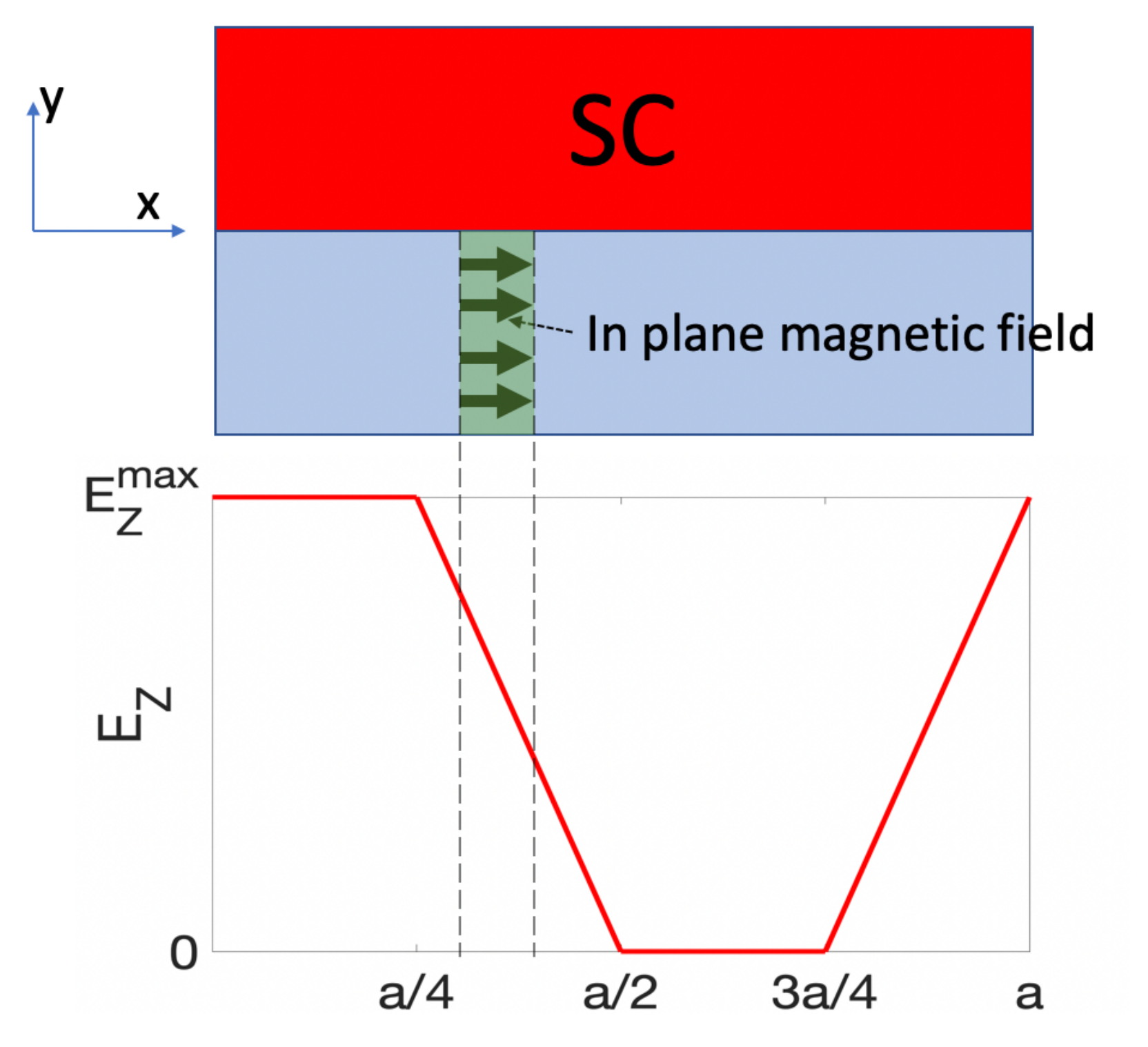}
\put(-115,110){(b)}
\end{minipage}
\label{8b}
}
\caption{(Color online). a: A schematic plot of the system. The superconducting order is induced only on the top half of the torus. An in-plane magnetic field is is confined in the domain wall and assumed to affect the bottom half of the torus; b: (top)The rectangular representation of the torus. The green shaded region is subject to in-plane magnetic field and is located near one of the domain walls. (bottom) The profile of the Zeeman coupling caused by the component of magnetic field in $z-direction$, which is perpendicular to the plane of the rectangular. Domain walls form in two regions [a/4, a/2] and [3a/4, a]. \label{fig8}}
\end{figure}

The Hamiltonian of the system is given by
\begin{eqnarray}
\label{e35}
\mathcal{H}=\mathcal{H}_t+H_{sc}+H_{bx}-\mu\hat{N}+C(\hat{N}-\hat{N}_0)^2.
\end{eqnarray}
The first term $\mathcal{H}_t$ is given by Eq.(\ref{e20}). As illustrated in Sec.II and IV, two domain walls form in the boundary regions between the spin polarized and unpolarized states. The boundary regions are the intermediate regions between $E_Z^{max}$ and 0 in Fig.\ref{8b}, which occur in the intervals [a/4, a/2] and [3a/4, a]. Each domain wall supports two counter-propagating edge modes with opposite spin polarizations. The second term $H_{sc}$ is the superconducting pairing term
\begin{eqnarray}
\label{e36}
H_{sc}=\int d\mathbf{ r}(\Delta(\mathbf{r})\Psi^\dagger_\uparrow(\mathbf{r})\Psi^\dagger_\downarrow(\mathbf{r})+\Delta^*(\mathbf{r})\Psi_\downarrow(\mathbf{r})\Psi_\uparrow(\mathbf{r})),
\end{eqnarray}
where the $\Delta(\mathbf{r})$ equals constant value $\Delta_1$ on the top half of the torus, and constant value $\Delta_2$ on the bottom half of the torus. In our simulations, $\Delta_2=0$, see Fig.\ref{8a}.
If we express the field operators $\Psi(\mathbf{r})$ in terms of the creation and annihilation operators $a^\dagger_j$ and $a_j$ that add or annihilate an electron in states given by Eq.(\ref{e19}), the superconductong pairing becomes $H_{sc}=\sum_{j,n} \Delta_{jn}a^\dagger_{j\uparrow}a^\dagger_{n\downarrow}+H.c.$, with $j,n=1,2,...,m$. When the total number of states $m$ is an even number, we obtain for $j+n=m,2m$
\begin{eqnarray}
\label{e37}
\Delta_{jn}&=&\sum_{k+q=-1}\frac{\Delta_1+\Delta_2}{2\sqrt{\pi}}\int_0^a dx \exp(-[\frac{(X_j+ka-x)^2}{2}\nonumber\\&+&\frac{(X_n+qa-x)^2}{2}]),
\end{eqnarray}
where $X_j=a\frac{j}{m}$. For $j+n$ odd numbers, we obtain:
\begin{eqnarray}
\label{e38}
\Delta_{jn}&=&\sum_{k,q}\frac{i(\Delta_2-\Delta_1)}{2\pi^{\frac{3}{2}}m(k+q+\frac{j+n}{m})}\int_0^a dx \exp\nonumber\\&(&-[\frac{(X_j+ka-x)^2}{2}+\frac{(X_n+qa-x)^2}{2}]).
\end{eqnarray}
The third term $H_{bx}$ is a spin-flipping tunneling term. In this section, this is the i- plane Zeeman coupling along the $x$-axis ($x$ and $y$ directions on the torus are defined in Fig.\ref{8b}). 
 In an in-plane magnetoc field, $H_{bx}=\sum_i\frac{1}{2}g\mu B(\mathbf{r_i})\sigma_x^{(i)}$. In our numerical calculations, $B(\mathbf{r})=B$ if $x\in[0.35a, 0.45a]$ and $y\in[0, 0.5a]$, where $a$ is the length of the torus in x- and y- directions. Otherwise $B(\mathbf{r})=0$, as shown in Fig.\ref{8b}. In the second quantization representation, $H_{bx}=\sum_{j,n} B_{jn}a^\dagger_{j\uparrow}a_{n\downarrow}+h.c.$, with $j,n=1,2,...,m$. For $j=n$, we have:
\begin{eqnarray}
\label{e39}
B_{jn}&=&\sum_{k=q}\frac{g\mu B}{4\sqrt{\pi}}\int_{0.35a}^{0.45a} dx \exp(-[\frac{(X_j+ka-x)^2}{2}\nonumber\\&+&\frac{(X_n+qa-x)^2}{2}]).
\end{eqnarray}
For the difference $j-n$ being odd numbers, the 
$B_{jn}$ is given by
\begin{eqnarray}
\label{e40}
B_{jn}&=&\sum_{k,q}i\frac{g\mu B}{2\pi^{\frac{3}{2}}m(k-q+\frac{j-n}{m})}\int_{0.35a}^{0.45a} dx \exp\nonumber\\&(&-[\frac{(X_j+ka-x)^2}{2}+\frac{(X_n+qa-x)^2}{2}]).
\end{eqnarray}
The fourth term in Eq.~(\ref{e35}) is the chemical potential, and the fifth term is the charging energy similar to that introduced in \cite{repellin2018numerical}. It represents the capacitor energy associated with the change of the number of electrons. These two terms are used to tune the electron number in the ground state of the system to the desired number. 

We now consider the Hilbert space of the numerical simulations. Here we take minimal possible number of  four electrons in the six orbitals defined in Eq.(\ref{e19}), so $m=6$. Two electrons (half of the total number) have the same spin, representing spin-polarized phase, and other two electrons represent the other half in spin-unpolarized state and have opposite spins. Thus, three electrons have spin up, and one electron is in spin down state, so the total spin of electrons $S=1$. We use the pair (N, S) to represent the set of states with total electron number N and total spin S. Without superconductivity and tunneling, the Hilbert space is (4, 1). The superconducting term $H_{SC}$ mixes the states with different numbers $N$, and the spin-flipping term $H_{BX}$ mixes the states with different total spins S, therefore our Hilbert space in numerical calculations is the following set of pairs\{(6, 2), (6, 1), (6, 0), (4, 2), (4, 1), (4, 0), (2, 1), (2, 0)\}. 

\begin{figure}
\centering
\subfigure
{
\begin{minipage}[b]{0.22\textwidth}
\includegraphics[width=4.5cm,height=4cm]{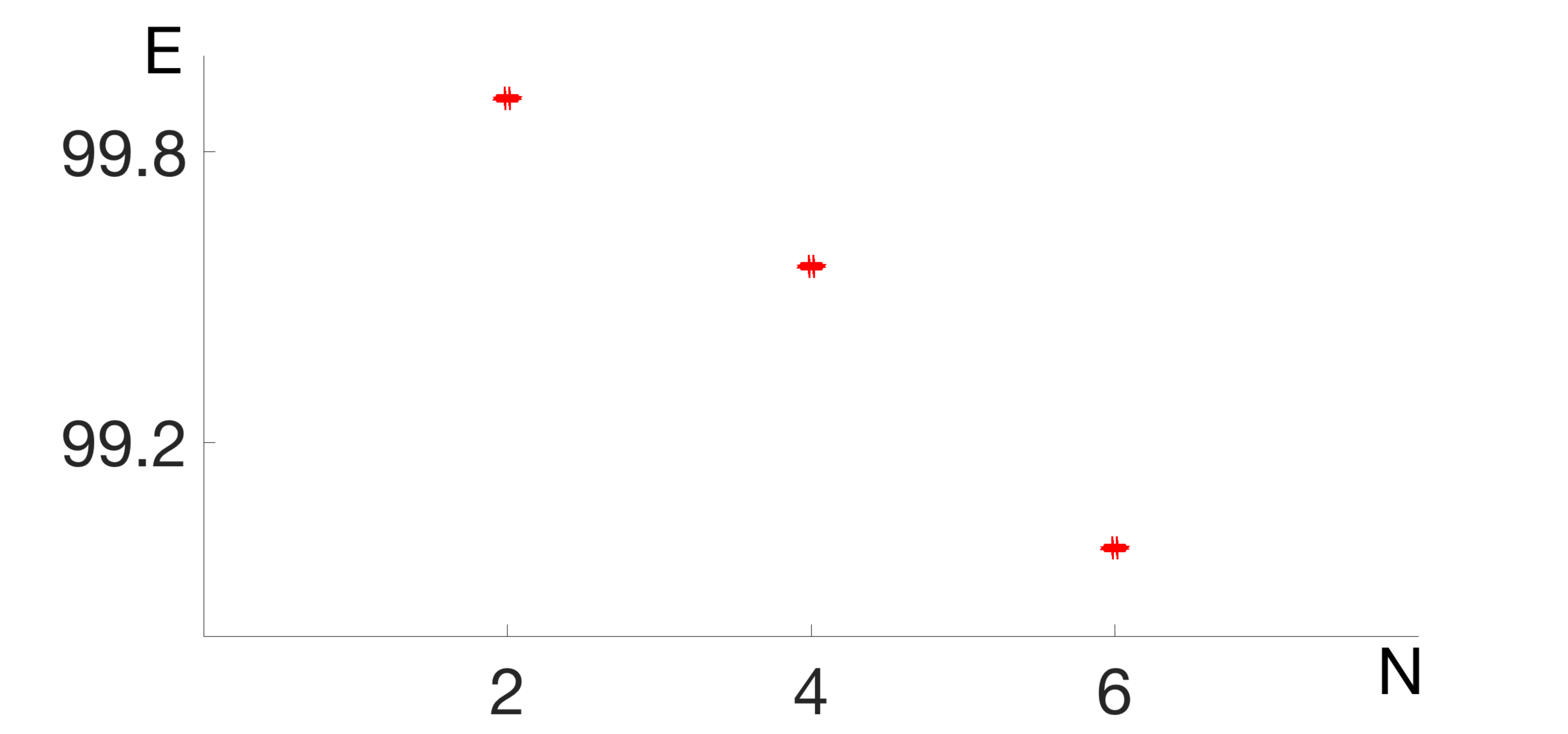}
\put(-130,110){(a)}
\end{minipage}

\label{9a}
}
\hfill
\subfigure
{
\begin{minipage}[b]{0.22\textwidth}
\includegraphics[width=3.8cm,height=4cm]{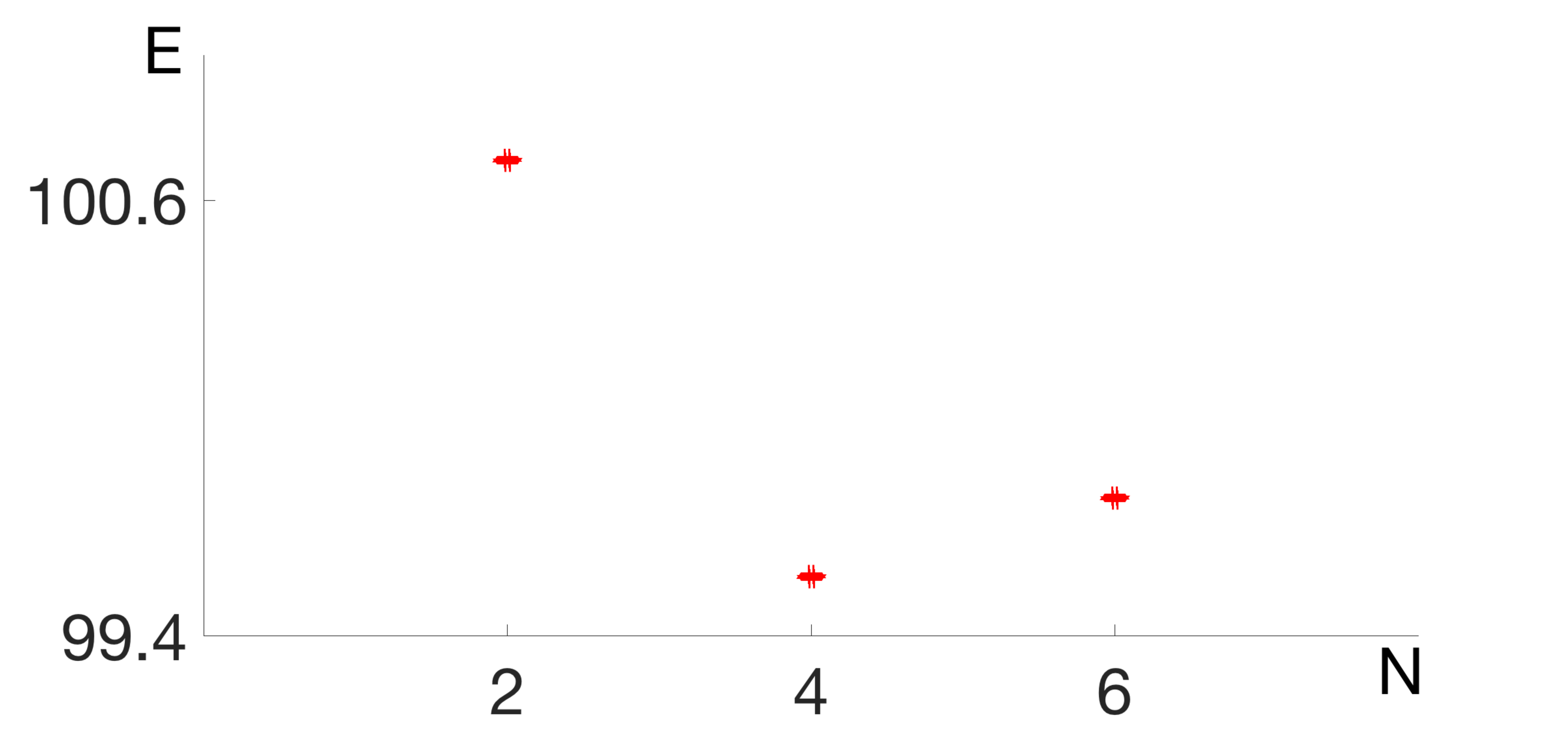}
\put(-115,110){(b)}
\end{minipage}
\label{9b}
}
\caption{(Color online). a: The lowest energies in $N=2,4,6$ sectors of $\mathcal{H}_t$. b: Including $\mu=0$, $C=0.2$, the lowest energies has been shifted so $N=4$ sector has the lowest energy and it's in the (4,1) sector. \label{fig9}}
\end{figure}

If Hamiltonian contains only $\mathcal{H}_t$, the lowest energies of the $N=2,4,6$ sectors are shown in Fig.\ref{9a}. The $N=6$ sector has a lower energy than $N=4$ sector. However choosing $\mu=0$, $C=0.2$, we find that the lowest energy is in the $N=4$ sector, see Fig.\ref{9b}. The lowest energy state is in the (4, 1) sector because of our special choice of the profile of the Zeeman coupling in z-direction (See Fig. \ref{8b}), guaranteing that (4, 1) states are more stable than (4, 2) and (4, 0) states. There are other ways to choose $\mu$ and $C$ in order to make (4, 1) the lowest energy states. We choose this special set because the half width of the BCS wave function is of the order of $\sqrt{N}$\cite{de2018superconductivity} so $N=2, 4, 6$ sectors all play important roles in the ground state properties. Therefore, a change of $\mu$ and $C$ will not affect the topological properties of the system. Experimentally $C$ should be a fixed number for the system and we only need to tune the chemical potential $\mu$. 

\begin{figure}
\centering
\subfigure
{
\begin{minipage}[b]{0.46\textwidth}
\includegraphics[width=8.5cm,height=4cm]{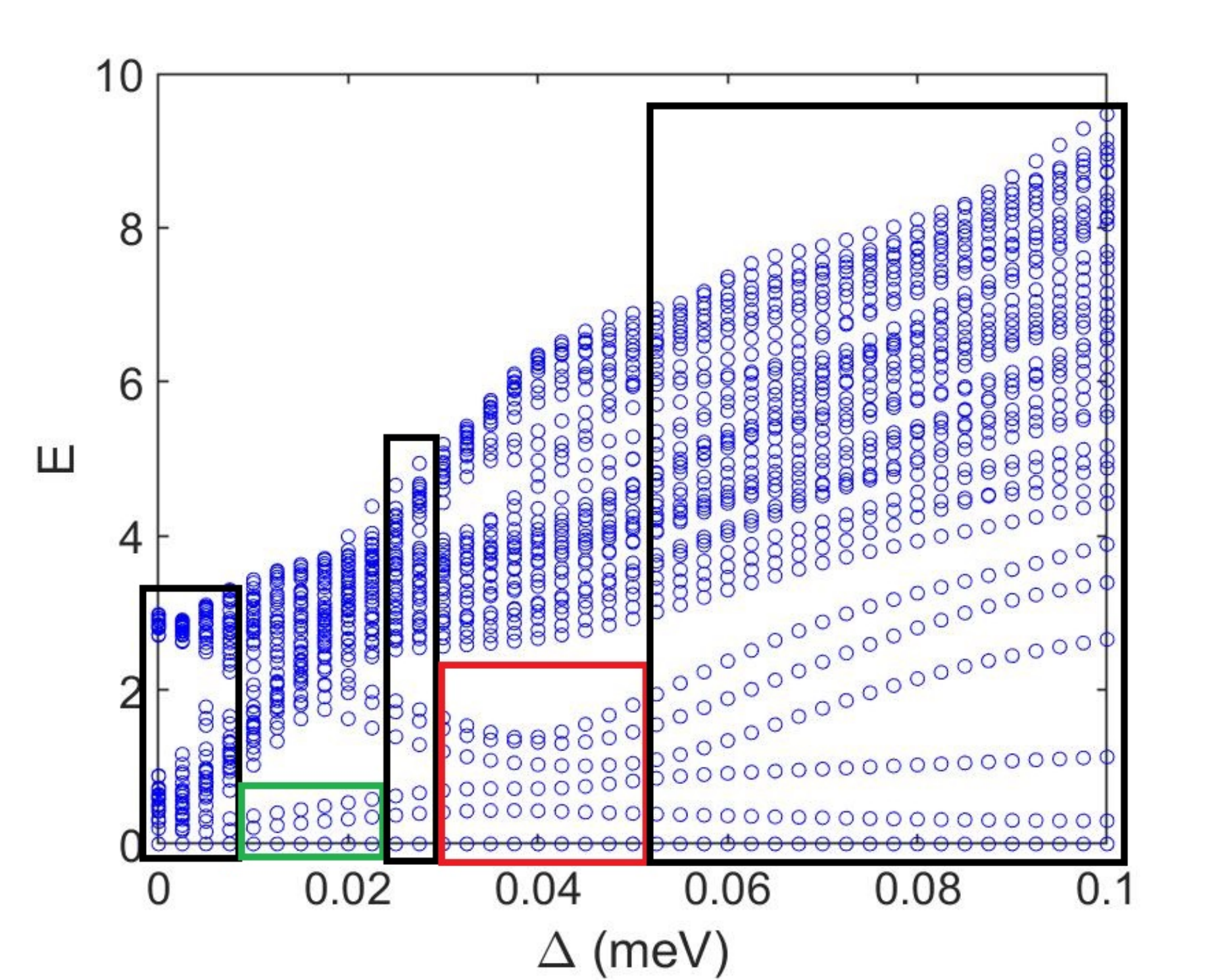}
\put(-232,110){(a)}
\end{minipage}

\label{10a}
}

\subfigure
{
\begin{minipage}[b]{0.46\textwidth}
\includegraphics[width=8.5cm,height=5cm]{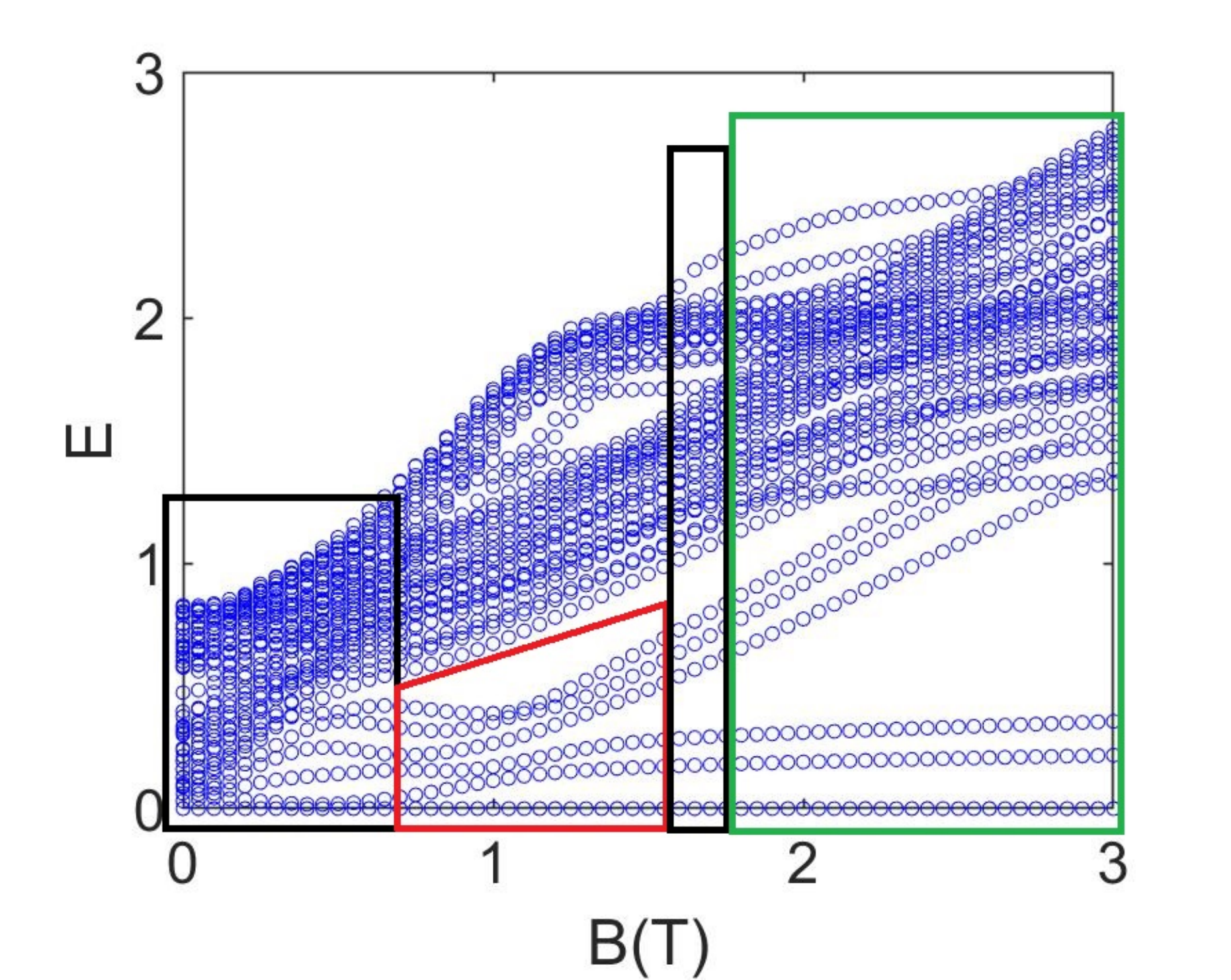}
\put(-232,110){(b)}
\end{minipage}
\label{10b}
}
\caption{(Color online). a: The energy dependence on superconducting pairing potential $\Delta$ with a fixed $B=1T$.  Red  rectangles indicate the range of parameters for the six fold degenerate ground state sub-space, which is separated from the bulk by a gap. This is the evidence for the appearance of parafermion modes. Green reactngle corresponds to a region, in which degeneracy  tends to three-fold. b: The energy dependence on the in plane magnetic field with a fixed $\Delta=0.05$meV. The six fold ground state degeneracy also appears and persists for a broader parameter regime. The energy is measured in units of $\frac{e^2}{\epsilon l_B}$  \label{fig10}}
\end{figure}

Now we include the $H_{sc}$ and $H_{bx}$ into the simulations. The special choice of a localized $H_{bx}$ allows us to focus only on a single domain wall. The edge states on the other domain wall will be gapped out due to the proximity superconducting order. In our system, the emergence of the parafermion mode means the appearance of a six-fold ground state degeneracy. Exactly diagonalizing Hamiltonian Eq.(\ref{e35}), we obtain the spectra shown in Fig.\ref{fig10}. In Fig. \ref{10a}, we fixed the value of B and change the superconducting pairing $\Delta$. We find that the system evolve from a single ground state to a three-fold ground state, and finally to a six-fold ground state. Based on general consideration of Sec. V, we assume that this six-fold ground state degeneracy represents the emergence of parafermions. The six states do not have exactly the same energy like in section 5. The reason is, arguments of section V apply, strictly speaking, for 1D systems, while our simulation treats a 2D system. Hence the degeneracy is lifted because of a possible tunneling between the edge states and other bulk orbitals. We observe that when the pairing potential is further increased, the system evolves into a three-fold degenerate state. The reason for this effect can be the system entering a gapped phase dominated by $\Delta$. In Fig. \ref{10b}, we fix the value of the order parameter $\Delta$ change the in-plane field $B$. We observe that the six-fold degenerate ground states appear at experimentally feasible $B$ of a few Tesla. When
$ B$ is increased even further, the system enters a tunneling dominated gapped regime with three-fold degenerate ground state. Thus, by exact diagonalization, we find that the system can enter a phase which has six-fold degenerate ground state. In our calculation with six particles, charging energy  restriction made the relevant values of $\Delta$  and $B$ quite experimentally feasible. The calculation, of course, must be extended to systems with larger number of particles in order to confirm this result.

\begin{figure}
\centering
\includegraphics[width=7.5cm,height=4cm]{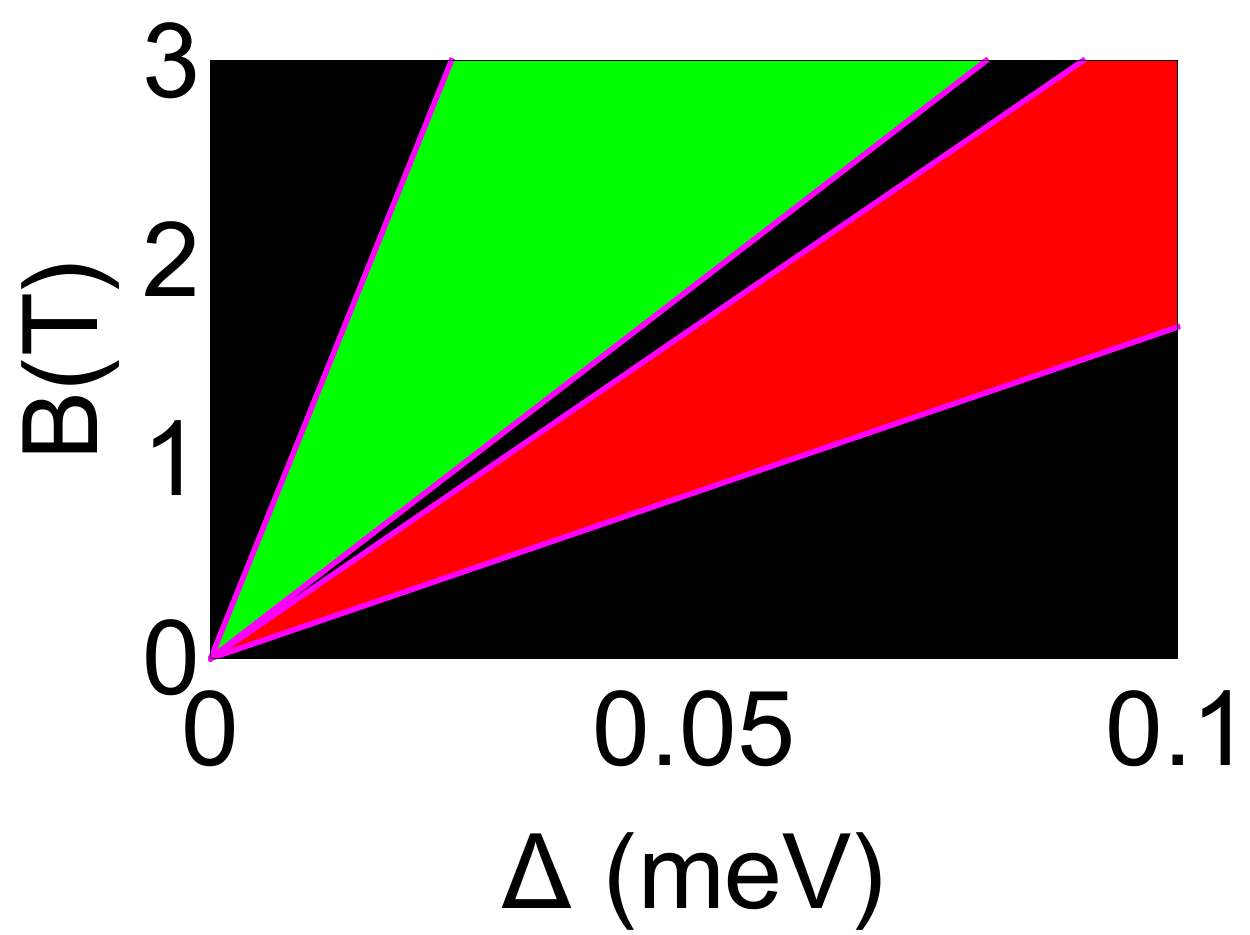}
\caption{(Color online). The phase diagram of our system. The red region represents states which has six fold ground state degeneracy. The green region represents states which has three fold ground state degeneracy. Black region represents gapless states. We identify a gap, when the maximum energy difference between candidate states should be at least two times as large as the second maximum energy difference. In this phase diagram we observe that the six fold ground state degeneracy regime are separated from other gapped states by gapless regions, which means that a quantum phase transition may occur between these regimes. We identifyl the phase represented by the red region as the topological superconducting phase supporting parafermion zero modes. \label{fig11}}
\end{figure}

To analyze the properties of the system further, we plot the phase diagram of the system in a wide range of $\Delta$ and $B$ in Fig.\ref{fig11}. We find that the phase A, which has six fold ground state degeneracy (represented by red  in Fig.\ref{fig11}), is separated from other gapped phases by a gapless incompressible regime. When we go from other gapped phases to phase A, the gaps first close and then reopen. A quantum phase transition may occur during this process. Combining numerical results with the analytic consideration, we conclude that it is legitimate to call the phase A the topological superconducting phase that supports parafermions.

\section{VII. Parafermion settings based on $\nu=\frac{4}{3}$ and $\nu=\frac{5}{3}$  spin transitions}.
Spin transitions in the fractional quantum Hall effect can also be observed at filling factors $\nu=\frac{4}{3}$ and $\nu=\frac{5}{3}$. In experiments on CdMnTe such a prominent transition at  $\nu=\frac{5}{3}$ was observed  in \cite{Weiss}. CdMnTe experiment at  $\nu=\frac{4}{3}$, although less prominently, also suggests a possible spin transition. 
These transitions can be easily understood both in terms of electron Landau level and compposite fermion Landau level pictures.

In terms of electron Landau levels, at low but quantizing magnetic fields, the splitting between spin-resolved Landau levels is dominated  by a large positive contribution of s-d exchange between electrons and Mn ions. The ground orbital Landau level in CdMnTe in magnetic fields about 3T is the ground orbital Landau level with spin down $(0, \downarrow)_e$ , and the next partially filled electron Landau level (1/3 or 2/3  filled for $\nu=\frac{4}{3}$ and $\nu=\frac{5}{3}$, correspondingly)  is the first excited orbital Landau level with the same spin, $(1, \downarrow)_e$ . While a positive exchange contribution to Zeemann splitting $E_{ex}$ becomes saturated at low magnetic fields and is independent of the magnetic field, the bulk negative g-factor leads to decrease of the Zeemann splitting $\hbar\Omega_Z$ with an increase of the  magnetic field. At the magnetic field such that $\hbar\omega_e - \hbar\omega_Z-E_{ex}=0$, levels $(0, \uparrow)_e$ and $(1, \downarrow)_e$  would cross in the absence of spin-orbit interactions. Instead, spin-orbit interaction leads to an anticrossing of Landau levels and anticrossing of the corresponding edge states \cite{kazakov2016,kazakov2017mesoscopic,simion2017disorder}. Nevertheless, at magnetic fields higher than the anticrossing field, the electron Landau level that is predominantly $(0, \uparrow)_e$ becomes partially filled.
Due to this transition the total electron spin polarization decreases from $5/3=1+2/3$ to $1/3=1-2/3$ for $\nu=\frac{5}{3}$, and from $4/3=1+1/3$ to $2/3=1-1/3$ for $\nu=\frac{4}{3}$, while the spin polarization of electrons in the first excited electron Landau level changes sign.

While this electron picture does not take into account electron-electron interactions, it is important for understanding these transitions, because it clearly shows that the transition involves anticrossing of levels and can be controlled via spin-orbit interactions. However, in order to understand the edge states of the system, we turn our attention to consideration of the composite fermion picture. 

The $\nu=\frac{4}{3}$ fractional quantum Hall state is a particle-hole symmetric to the principal states, $4/3= 2- 2/(4-1)$ and can be described by two filled composite fermion Landau levels. In quantizing but sufficiently low magnetic fields in CdMnTe, these two filled CF hole levels are  $(0, \downarrow)_{cf}$ and  $(1, \downarrow)_{cf}$, representing a spin-polarized CF phase. At sufficiently high magnetic fields, the two filled levels are  
$(0, \downarrow)_{cf}$ and $(0, \uparrow)_{cf}$, and composite fermions become spin-unpolarized. At the true boundary of the sample, the edge starts with bulk density 4/3, raises to density 2, decreases to density 1 and then goes to 0 \cite{beenakker1990edge}. The edge corresponding to the filled ground level $\nu=1$ goes around the sample, and can be removed from low energy theory from the internal boundary between two differently spin-polarized regions.
At this boundary, for one of the phases we have two hole composite fermion edges  responding to $(0, \downarrow)_{cf}$ and  $(1, \downarrow)_{cf}$ Lambda- levels  closer to the bulk of the corresponding region, followed by 
the outer edge corresponding to the electron Landau level $(1, \downarrow)_e$. For the other phase region, the internal CF hole edges correspond to  $(0, \downarrow)_{cf}$ and $(0, \uparrow)_{cf}$ $\Lambda$ levels, and the outer edge corresponding to electron Landau level $(0, \uparrow)_e$. The question now arises whether any of the edges can be removed from low energy picture. Generally 1/3 charge quasiparticles cannot tunnel through the region of charge 1 quasiparticles, and that is a correct for the true edge of the sample. However, the closest edge states in the domain wall originating from the  Landau levels  $(1, \downarrow)_e$ and  $(0, \uparrow)_e$ become hybridized and form a helical domain wall, which conducts in the presence of impurities or smooth random potential and constitutes a compressible region \cite{simion2017disorder}. Hence there is no longer a prohibition for 1/3 charge channel to tunnel and couple to another 1/3 charge through this compressible channel. Then, by using the argument similar to that in Sec. II, we can remove  edge states corresponding to $(0, \downarrow)_{cf}$ in both phases from the low-energy sector. The remaining $(0, \uparrow)_{cf}$ and $(1, \downarrow)_{cf}$ states constitute two counterpropagating hole charge 1/3 states 
with opposite spin. Interestingly enough, they coexist with a helical domain wall made of  hybridized $(1, \downarrow)_e$ and  $(0, \uparrow)_e$ electron channels. However, the superconducting coupling and spin-flipping interaction can still produce gaps for1/3 charge counterpropagating states with opposite spins, leading to parafermions. At the same time, a helical domain wall made of hybridized $(1, \downarrow)_e$ and  $(0, \uparrow)_e$ channels will result in Majrorana modes \cite{simion2017disorder}.

 The $\nu=\frac{5}{3}$ fractional quantum Hall state and spin transition is desribed similarly. $5/3= 2- 1/(2+1)$ and can be described by one filled composite fermion hole Landau level. In weaker magnetic fields, this is the $(0, \downarrow)_{cf}$ level, and in higher magetic fields, this is the $(0, \uparrow)_{cf}$ hole level. However, analysis shows that at the domain wall boundary between two phases with diffeernt spin polarization, these two CF edge states are again separated by a helical domain wall made of hybridized $(1, \downarrow)_e$ and  $(0, \uparrow)_e$ electron channels. Similarly to $\nu=\frac{4}{3}$ domain wall, $\nu=\frac{5}{3}$domain wall suggests coexistense of parafermions and Majorana fermions. The structure of the domain wall and coexisting  $(1, \downarrow)_e$ and  $(0, \uparrow)_e$s edges and  $(0, \downarrow)_{cf}$ $(1, \uparrow)_{cf}$  edges.

\begin{figure}
\includegraphics[width=8.7cm,height=5.4cm]{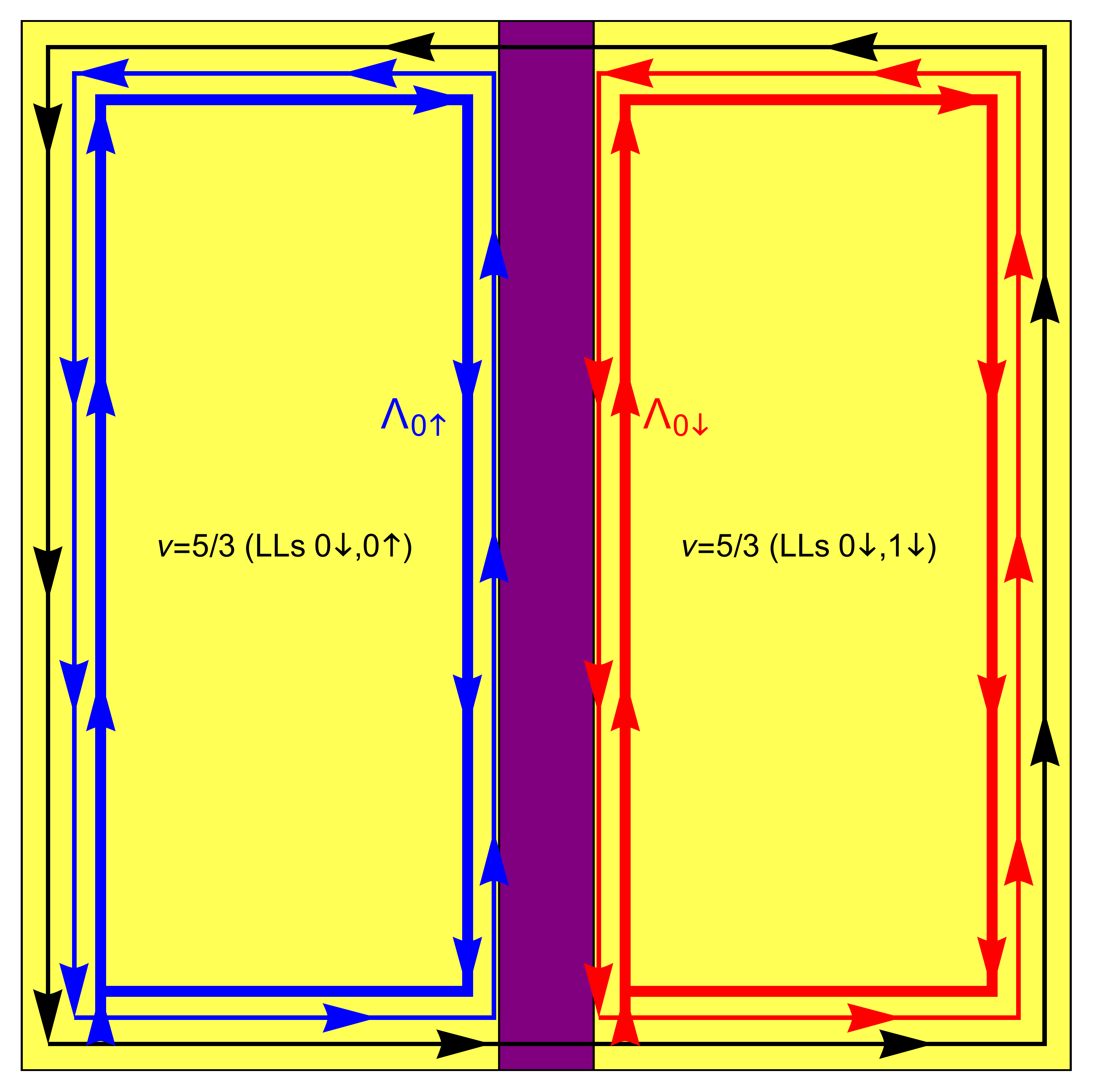}
\caption{(Color online). The domain wall and edge states at $\nu=\frac{5}{3}$.Electrons occupy  $(0, \downarrow)_e$ and  $(0, \uparrow)_e$ Landau levels in a phase with total spin density 5/3 
and $(0, \downarrow)_e$ and  $(1, \downarrow)_e$ Landau levels in a phase with total spin density 1/3. The outermost black  edge channel around both regions is integer $(0, \downarrow)_e$  edge. Closest edges in the domain wall area are edges corresponding to integer levels  $(0, \uparrow)_e$ ( blue) and  $(1, \downarrow)_e$  (red). The innermost levels are composite fermion hole $\Lambda$-levels  $(0, \downarrow)_{cf}$ (red) $(1, \uparrow)_{cf}$  (blue) \label{Fig53}}
\end{figure}

An interesting question emerges whether the spin-flip gapping mechanism for the $(0, \uparrow)_{cf}$ and $(1, \downarrow)_{cf}$ counterpropagating CF hole states, can be stronger due to spin-orbit interactions,  in contrast to similar CF states discussed in Sec. II.  For $\nu=\frac{2}{3}$ states, composite fermion states are produced by default from electron states of the ground Landau level. If our starting point is these ground landau level states, and we consider spin-orbit interactions as a perturbation, then all intra-Landau level spin-orbit matrix elements vanish, and spin-orbit does not have any effect  in the bulk. In the appendix, we demontrate that for the domain wall that defines a mid-plateau peak in the resistivity, the edge states of $(0, \uparrow)_{cf}$ and $(1, \downarrow)_{cf}$  composite fermions levels are characterized by an extremely small spin-orbit gap in the bulk. 

The $\nu=\frac{4}{3}$ states in CdMnTe originate from full $(0, \downarrow)_e$  Landau  level and $(0, \uparrow)_e$ Landau level in the phase with 2/3 total spin polarization, 
and from full $(0, \downarrow)_e$  Landau  level and from $(1, \downarrow)_e$ Landau level in the phase with 4/3 total spin polarization. In the latter case, if we include $(1, \downarrow)_e$ level into a starting point for composite fermions, we clearly go beyond lowest Landau level approach to composite fermions. However, large exchange splitting that is nearly independent of magnetic feild  in CdMnTe makes the $(1, \downarrow)_e$ Landau level much lower in energy than  $(0, \uparrow)_e$ state in a wide range of magnetic fields. Lowest Landau level restriction is justified, of course, in the limit of very large magnetic fields. However, spin transitions, crossing and anticrossing of levels take place at much smaller fields, and their consideration with inclusion of higher electron Landau levels is justified. Once the $(1, \downarrow)_e$ level is included in the Hilbert space for evaluating interactions, the spin-orbit effects become important.

 To illustrate the importance of spin-orbit effects, it is also worth emphasising that when spin-orbit effects are sizable, it is of interest to consider the problem explicitly taking into account spin-orbit interactions right from the beginning. CdMnTe quantum wells are 
affected by the spin-orbit interactions of Rashba type\cite{Rashba1960}, which can be described by the spin-dependent vector potential\cite{LGM,ILGP,SLG}, which acts together with electromagnetic vector potential defined by the magnetic field. In a perpendicular magnetic field ${\mathbf H}=(0,0, H)$ the electron Hamiltonian is given by
\begin{equation}
\label{LR}
{\cal H}= \frac{1}{2m^*}\left({\mathbf p}-\frac{e}{c}{\mathbf A} +{\mathbf A}_{so}  \right)^2  + \frac{1}{2}\left( g\mu H  +E_{ex}\right){\mathbf \sigma}_z,
\end{equation}
where ${\mathbf A}$ is the magnetic vector-potential, the spin-dependent vector potential ${\mathbf A}_{so}=\alpha m ({\mathbf \sigma}_y, -{\mathbf \sigma}_x,0)$, $m^*$ and $g$ is the effective electronmass and g-factor, orrespondingly, and $ E_{ex}$ is the contibution of Mn ions in the spin-splitting of electron levels taken in the mean field approximation. We assume the Rashba constant to be spatially independent and include constant energy term 
$ \alpha^2 m $i nto electron Hamiltonian. The energy eigenvalues for this problem are given by
\begin{eqnarray}
& E_0=\frac{1}{2}\left( \hbar\omega_c + g\mu H  +E_{ex}\right)+ \alpha^2 m\\
&E_{n,\pm}=\hbar\omega_c n+ \alpha^2 m\nonumber\\
&\pm \sqrt{\frac{1}{4}\left(g\mu H+E_{ex} +\hbar\omega_c\right)^2+\left(\hbar\alpha/\ell_m\right)^2n},
\end{eqnarray}
The eigenfunctions have the following form:
\begin{eqnarray}
& \psi_0=\left(
\begin{array}{c}
u_0\\
0
\end{array}\right); \hspace{3mm}
& \psi_{n,\pm}=\left(
\begin{array}{c}
a_{n,\pm}u_n\\
b_{n,\pm}u_{n-1},
\end{array}\right),
\end{eqnarray}
where spinor coefficients $a_{n,\pm}$ and $b_{n,\pm}$ are nonzero,and $\vert a_{n,\pm}\vert^2+\vert b_{n,\pm}\vert^2=1$

For electrons in CdMnTe at small magnetic fields, due to large $E_{ex}$, the ground state energy is $E_{1,-}$, and the first excited level has energy $E_{2,-}$. The electron spectrum is characterized both by crossings, e.g.,of $E_0$ and $ E_{1,-}$ levels, and by anticrossings, e.g., of $E_{2,-}$ and $E_0$ levels. Is is noteworthy that spin in the ground level at small fields deviatets from $z-$direction due to spin-orbit interactions,  hence it is certainly important for composite particles in this regime.  

We now include spin-orbit interactions in the Chern-Simons procedure. The mean-field Hamiltonian of the system reads
\begin{equation}
\label{CS}
{\cal H}= \frac{1}{2m^*}\left({\mathbf p}-\frac{e}{c}{\mathbf A}^* +{\mathbf A}_{so}  \right)^2  + \frac{1}{2}\left( g\mu H  +E_{ex}\right){\mathbf \sigma}_z +V,
\end{equation}
where ${\mathbf A}^*$ is the effective vector potential that takes into account both external magnetic and average Chern-Simons magnetic field, and V are electron-electron interactions.
The naive solution for energies of composite fermion levels becomes
\begin{eqnarray}
& E_0=\frac{1}{2}\left( \hbar\omega^*_c + g\mu H  +E_{ex}\right)+ \alpha^2 m\\
&E_{n,\pm}=\hbar\omega^*_c n+ \alpha^2 m\nonumber\\
&\pm \sqrt{\frac{1}{4}\left(g\mu H+E_{ex} +\hbar\omega^*_c\right)^2+\left(\hbar\alpha/\ell^*_m\right)^2n},
\end{eqnarray}
Thus, in the naive solution, in addition to the effective magnetic field defining the cyclotron frequency, the spin-orbit term is also expressed in terms of magnetic length characterizing the effective rather than the external magnetic field. It is, of course clear, that this spectrum has to be renormalized by interactions. Nevertheless, it is apparent that it can be characterized by both crossings and anticrossings. Edge channels originating from CF levels that experience antiicrossing can be strongly gapped by spin-orbit interactions. 
 
\section{VIII. Summary}
In this paper, we have considered first the domain wall between spin-polarized and spin-unpolarized regions of the 2D electron liquid in $\nu=\frac{2}{3}$ fractional quantum Hall state.  Analytic considerations show that the domain wall between the spin-polarized and spin-unpolarized regions of the 2D electron liquid  hosts two edge states that are counter-propagating and have opposite spin polarization. This picture is confirmed using numerical calculations in a disk and torus geometries. We proved both analytically and numerically that the edge states can support parafermions when the domain wall area is proximity-coupled to an s-wave superconductor. Hilbert space for exact digonalization study significantly increases due to account of superconducting correlations. We have discussed control of parafermion zero modes due to hybridization of edge states by spin-flipping interactions. In GaAs for $\nu=\frac{2}{3}$ spin transitions, a tilted magnetic field with an in-plane component controlling spins can be used for gapping edge channels, while spin-orbit interactions are negligible. In $\nu=\frac{4}{3}$ and In $\nu=\frac{5}{3}$ spin transitions  in fractional quantum Hall effect in $CdMnTe$, parafermions modes also emerge and can be controlled by electrostatic gates due to sizable spin-orbit anticrossing gap. We discuss near absence of spin-orbit coupling for composite fermions at  $\nu=\frac{2}{3}$ for all principal composite fermion states, and emergence of spin-orbit interactions in systems like  CdMnTe in the presence of exchange splitting of electron states. In these systems, spin-orbit interactions arise for states that are particle-hole symmetric to the principal states in the presence of spin, such as $\nu=\frac{4}{3}$ and In $\nu=\frac{5}{3}$.

\section{Acknowledgement}
Authors thank Martin Kruczenski and Bin Yan for helpful discussions. This work is supported by the U.S. Department of Energy, Office of Basic Energy Sciences, Division of Materials Sciences and Engineering under Award DE-SC0010544 (Y.L-G, G.S. and J.L.). Modeling of the spin-polarized and unpolarized phases on the disk was supported by the Department of Defense Office of Navel Research Award N000141410339 ( G.S and Y.L-G). 

\appendix*
\section{Spin-flipping interactions in the lowest Landau level.}

Tunneling between two counterpropagating fractional quantum Hall edge states with opposite spin polarization requires spin-flip interactions. One candidate is spin-orbit interaction suggested for setting   
considered in \cite{clarke2013exotic}, and the other is spin-flip due to an in-plane component of a tilted magnetic field. We now analyze the viability of these interactions for generating tunneling gap.

The crossing in quasiparticle spectra in our system occurs between composite fermion levels. In consideration at the level of composite fermions, the "orbital quantization" energy, or cyclotron frequency is entirely determined by the  intra lowest  Landau level electron-electron processes. Landau level mixing can not change the picture as long as one always considers electrons of the ground electron level as giving rise to composite fermion quasiparticles. For this reason, the dispersion of composite fermions is defined by the Coulomb energy, and the effective band mass of electron plays no role. Composite fermions couple to the effective magnetic field, defined by a compensation of the external magnetic field and the Chern-Simons field. In contrast, Zeeman coupling of composite fermions is described by the value of the electron g-factor and an external magnetic field.  

Considering composite fermions microscopically, one generally  has to start from spin up and down electrons of the ground Landau level and include electron-electron interactions. Spin-orbit interactions, such as the Rashba or Dresselhaus spin-orbit coupling can be included perturbatively. It is clear, however, that for the "bulk" 2D electrons, the matrix elements of the spin-orbit coupling between up and dow states with the same orbital wavefunction simply vanish. That is not the case for transitions between "edge" modes. Indeed, our setting includes either two electrostatic gates with differing voltages, leading to different electron densities underneath them and ultimately to composite fermion level structure shown in Fig 2a, or can be described by the varying electron Zeeman coupling leading to the same picture. In both cases, electrons in the domain wall that eventually become composite fermion edge states are subject to a spin-dependent potential and the corresponding electric field that is opposite for the two electron spin directions.

Then the lowest Landau level wavefunctions  for two spin directions in the Landau gauge with magnetic field ${\mathbf H}\parallel z$ vector potential ${\mathbf A}=(0,Hx,0)$ and spin-dependent electric fields in 
$x-$direction with magnitude $E$ are given by
\begin{eqnarray}
\Psi_{\uparrow}&=& \exp{ik_yy} \frac
{\exp{\left[-\frac{1}{2}\left(\frac{x-x_{\uparrow}}{l_m}\right)^2\right]}}
{\sqrt{l_m}}\\
\Psi_{\downarrow}&=& \exp{ik_yy}\frac
{\exp{\left[-\frac{1}{2}\left(\frac{x-x_{\downarrow}}{l_m}\right)^2\right]}}
{\sqrt{l_m}},
\end{eqnarray}
where 
\begin{eqnarray}
x_{\uparrow}&=& -l_m^2 k_y - \frac{eEl_m^2}{\hbar\omega_c},\\
x_{\downarrow}&=&-l_m^2 k_y + \frac{eEl_m^2}{\hbar\omega_c},
\end{eqnarray}
 $\omega_c$ is the electron cyclotron frequency and $k_y$ is the electron wavevector along the domain wall.
$\Psi$'s are the wavefunctions in the corresponding crossed electric and magnetic fields. These two wavefunctions are not orthogonal due to different  $x_{\uparrow}$ and $x_{\downarrow}$.
The matrix element of the, e.g., Rashba spin-orbit interaction described by the Hamiltonian $H_R=\alpha (\sigma_x k_y -\sigma_y k_x)$, with $k_x=-i\frac{\partial}{\partial x}$, is given by
\begin{equation}
\langle \Psi_{\uparrow}|H_R|\Psi_{\downarrow}\rangle= \alpha\frac{eE}{\hbar\omega_c}\exp{\left[-\left(\frac{eEl_m}{\hbar\omega_c}\right)^2\right]}.
\end{equation}
At electric field $E=0$ the spin-orbit matrix element vanishes, as it should be in the bulk 2D electrons case. The eq. A43 is a perturbative result obtained using the lowest Landau orbital wavefunctions, but it contains $\hbar\omega_c$ in the denominator as if it is related to an admixture of a higher Landau level. This is not accidental. Indeed using consideration of Rashba interaction in the previous section , the result A50  can then be obtained in the leading order by calculating the matrix element of the interaction describing the spin-dependent electric field, $ H_E= eEx\sigma_z$ between two exact bulk spin-orbit states.

We now consider the magnitude of the spin-orbit gap for edge modes, which is governed bythe magnitude of the electric field  $E$. 
The experimentally possible narrowest domain wall (in the $x-$direction) is approximately 50nm, and so that the magnitude of the gap is defined by the difference of energies of the like spins underneath the two electrostatic gates, Fig 2a. This value of this energy difference, however, is strongly restricted by the requirement
that the fractional quantum Hall system underneath both gates is within the $\nu=2/3$ quantum Hall plateau.  Such restriction means that the difference of energies on the left and right of the domain wall is limited by the Zeeman energy that corresponds to the length of the interval of magnetic fields corresponding to the plateau, i.e. the plateau width. In experiments  \cite{wu2017parafermion}, where the domain wall in the fractional quantum Hall effect was observed and electrostatically controlled, the measured width of the plateau is $\delta B=0.35 T$. In this case the magnitude of the spin-orbit matrix element in GaAs system is negligibly small $\sim 1-2 \mu eV$.

Therefore for the domain wall setting, experimentally feasible in-plane Zeeman field of order of 1T, that leads to a tunneling gap sufficient for the emergence of parafermions in our modeling, becomes a preferential mechanism for generating a hybridization of edge states with opposite spin and a corresponding tunneling gap. 

The situation changes, however, for a settings based on spin transitions at $\nu=4/3$ and $\nu=5/3$  in a system like CdMnTe \cite{Weiss} discussed in Sec. VII. In both of these cases,  edge states near a tentative spectral crossing belong to different electron Landau levels, and spin-orbit interactions result in anticrossing with an anticrossing gap in CdMnTe of the order to 50$\nu eV$ \cite{simion2017disorder}. This spin-orbit gap can be effectively controlled by the electrostatic gate, which then can control emerging parafermions similarly to control of Majorana zero modes in \cite{simion2017disorder}.

%

\end{document}